\documentclass[amsmath,amssymb,10pt,aps,twocolumn]{revtex4-2} 
\usepackage[colorlinks=true]{hyperref}
 \usepackage{lipsum}
 \usepackage{graphicx}
 \usepackage{textcomp}
\usepackage[dvipsnames]{xcolor}
\usepackage{latexsym}
\usepackage{amsmath}
\usepackage{amsthm}
\usepackage{amssymb}
\usepackage{epstopdf}
\usepackage{enumitem}
\usepackage{setspace} 
\usepackage{dcolumn} 
\usepackage{bm} 
\usepackage{setspace} 
\usepackage{slashed}
\usepackage{color}
\usepackage{xcolor}
\usepackage{youngtab}
\usepackage{tikz}
\usepackage{tikz-3dplot}
\usepackage{braket}
\usepackage{mhchem}

\usetikzlibrary{shapes,snakes,arrows,chains,matrix,positioning,scopes,calc}
\usetikzlibrary{decorations.markings}

\usepackage[tabbotcap]{subfigure}
\newcommand{\rev}[1]{{\color{red}#1}}

\allowdisplaybreaks

\usepackage{lipsum}

\newlength{\diamondrulelength}
\newlength{\diamondrulethickness}
\newcommand{\diamondrule}{\begin{center}\tikz{\fill[black] (0.5\diamondrulelength,0) -- (0,0.5\diamondrulethickness) -- (-0.5\diamondrulelength,0) -- (0,-0.5\diamondrulethickness) -- cycle;}\end{center}}

\begin{document}
\title{Quantum Spin Liquids in Weak Mott Insulators with a Spin-Orbit Coupling}

\author{Asimpunya Mitra} 
\email{asimpunya.mitra@mail.utoronto.ca}
\affiliation{Department of Physics, University of Toronto, Toronto, Ontario M5S 1A7, Canada}
\author{Daniel J. Schultz}
\email{d.schultz@mail.utoronto.ca}
\affiliation{Department of Physics, University of Toronto, Toronto, Ontario M5S 1A7, Canada}
\author{Yong Baek Kim}
\email{ybkim@physics.utoronto.ca}
\affiliation{Department of Physics, University of Toronto, Toronto, Ontario M5S 1A7, Canada}

\date{\today}   

\begin{abstract}
The weak Mott insulating regime of the triangular lattice Hubbard model exhibits a rich magnetic phase diagram as a result of the ring exchange interaction in the spin Hamiltonian. These phases include the Kalmeyer-Laughlin type chiral spin liquid (CSL) and a valence bond solid (VBS). A natural question arises regarding the robustness of these phases in the presence of a weak spin-orbit coupling (SOC). In this study, we derive the effective spin model for the spin-orbit coupled triangular lattice Hubbard model in the weak Mott insulting regime, including all SOC-mediated spin-bilinears and ring-exchange interactions. We then construct a simplified spin model keeping only the most relevant SOC-mediated spin interactions. Using infinite density matrix renormalization group (iDMRG) we show that the CSL and VBS phases of the triangular lattice Hubbard model can be stabilized in the presence of a weak SOC. The stabilization results from a compensation between the Dzyaloshinskii-Moriya interaction and a SOC-mediated ring-exchange interaction. We also provide additional qualitative arguments to intuitively understand the compensation mechanism in the iDMRG quantum phase diagrams. This mechanism for stabilization can potentially be useful for the experimental realization of quantum spin liquids.
\end{abstract}

\maketitle

\section{Introduction}

Quantum spin liquids are long-range entangled quantum ground states of interacting spin systems, which host fractionalized excitations and emergent gauge fields \cite{Savary_2016,nature_Leon_Balents,annurev-Knolle,RevModPhys_Wen,RevModPhys_Zhou,Science_Broholm}. Realization of such exotic quantum states in real materials has been a long-standing challenge in quantum materials research. Frustrated magnets are promising platforms for quantum spin liquids as they offer either geometric frustration from the lattice structure or the presence of competing interactions, which promote magnetic frustration and inhibit magnetic ordering. Various approaches for generating competing interactions have been proposed, ranging from the bond-dependent anisotropic interactions (e.g. Kitaev) \cite{Kitaev_2006,Takagi_2019,Banerjee_2016,Liu_2018,Liu_2020,Sano_2018}, to the inclusion of further neighbor spin interactions \cite{Halloran_2023,Das_2021,Bose_2023,Liu_2023}.

Another novel route to achieve frustration via competing interactions is to utilize the charge fluctuations in weak Mott insulators \cite{He_2018,Sheng_2009,Block_2011,Motrunich_2005,Kaneko_2014,Yang_2010,Mishmash_2015,Podolsky_2009,Senthil_2008,Senthil_2008b,Zou_2016,Tang_2023,Delannoy_2005,Grover_2010, Mishmash_2013,Hu_2015,Tocchio_2013,Zhu_2015}. In this case, the virtual charge fluctuations can lead to significant multi-spin interactions because of the small charge gap. For example, starting from the Hubbard model at half-filling, the strong coupling expansion in terms of $t/U$ \cite{t_U_Hubbard_MacDonald,SW_transformation}, where $t$ and $U$ are the hopping and on-site repulsion, leads to four-spin ring-exchange interactions that compete with the nearest and further neighbor two-spin interactions. The Hubbard model on the triangular lattice with a moderate charge gap \cite{Shirakawa_2017,Venderley_2019,Aghaei_2020,Szasz_2020_PRX,Chen_2022,Szasz_2021,Zhu_2022,Banerjee2023,Kuhlenkamp2024} and its effective spin models including ring-exchange interactions \cite{Cookmeyer_four_spin,schultz2023electric,Szasz_2020_PRX} have been extensively studied using DMRG. In the $J_1$-$J_{\text{ring}}$ model \cite{Cookmeyer_four_spin}, it was found that tuning the ratio of $J_{\text{ring}}/J_1$ between $0$-$0.4$ changes the ground state successively from a $120^{\circ}$ order, to a Kalmeyer-Laughlin chiral spin liquid (CSL) \cite{Laughlin_1987,Laughlin_1989}, valence bond solid (VBS), and finally into a zig-zag ordered state. The phase diagram of the $J_1$-$J_2$-$J_3$-$J_{\text{ring}}$ spin model \cite{schultz2023electric} also shows a similar structure when parameterized by $t/U$. 

Typically in weak Mott insulators, the local moments come from $4d$ or $5d$ atomic orbitals, so a significant spin-orbit coupling (SOC) may be present \cite{Krempa_2014}. Hence it is important to understand the effect of the SOC on quantum spin liquid phases that may be obtained via the ring-exchange or multi-spin exchange interactions in weak Mott insulators. Deep in the Mott insulating phase, two-spin interactions dominate due to a large charge gap, however, multi-spin interactions are of increasing importance closer to the Mott transition. The strong coupling expansion (in $t/U$) of the Hubbard model with a SOC leads to the addition of the Dzyaloshinskii-Moriya (DM) interaction at the leading order for inversion symmetry broken systems \cite{Dzyaloshinsky_1958,Moriya_1960,Spin_triplet_VBS,Krempa_2012}. The conventional wisdom is that the DM interaction (or the SOC) would lift the frustration and promote a magnetically ordered state. For example, on the triangular lattice, the DM interaction would favour a $120^{\circ}$ ordered ground state \cite{Hog_2022}. On the other hand, the presence of SOC in weak Mott insulators would lead to SOC-mediated multi-spin interactions (SOC-mediated ring exchange) along with the conventional ring exchange interaction. This can potentially lead to competition between the two opposing trends: increased frustration due to the ring-exchange interactions and decreased frustration due to the DM interaction. Hence, it is of considerable interest to investigate this problem, as this might lead to a route of stabilizing quantum spin liquids in the presence of a SOC.

In this work, we model weak Mott insulators in the presence of a spin-orbit coupling. We start from a single-band Hubbard model on the triangular lattice at half-filling with a SOC term in the Hamiltonian. Working in the strong-coupling limit, we derive an effective spin model with all the spin-bilinear and multi-spin exchange interactions. Apart from the Heisenberg and ring-exchange interactions, we find that the SOC generates the DM interactions and anisotropic symmetric exchange interactions, to leading order in $t/U$. Expanding to higher orders, we find new types of SOC-mediated ring-exchange interactions. 
Considering the strength of the SOC to be small compared to the electron hopping energy, we work with only the most relevant SOC-mediated exchange interactions in the Hamiltonian, which are the nearest-neighbour DM interaction and the leading order SOC-mediated ring exchange interaction. Using iDMRG \cite{White_1992,White_1993,RevModPhys_Schollwock,mcculloch_2008,Vidal_2008}, we construct a zero-temperature quantum phase diagram for this model. We demonstrate that the CSL and VBS phases are stabilized in the presence of a weak SOC in some parameter regimes. This stability is due to the SOC-mediated ring exchange term compensating for the DM interaction, either of which individually tends to induce a $120^{\circ}$ order, but with opposite handedness.

The rest of the paper is organized as follows. In Sec. \ref{sec:model} we introduce the Hubbard model with a SOC and describe the corresponding low-energy effective spin model. In Sec. \ref{sec:QPD} we present the quantum phase diagrams obtained using iDMRG in the presence of a weak SOC, and highlight the stability of the CSL and VBS phases. In Sec. \ref{sec:mft} we propose a qualitative argument to understand the presence and stability of CSL and VBS phases appearing in the phase diagrams. Finally, we provide a discussion in Sec. \ref{sec:disc}.

\section{Model}\label{sec:model} 
\subsection{Hubbard Model with a SOC}
We consider a single-band Hubbard model at half-filling on the triangular lattice, with a spin-orbit coupling term built into the hopping matrix $\Tilde{t}$ \cite{Spin_triplet_VBS,Krempa_2012}. The Hamiltonian for the model is
\begin{equation}
H = -\sum_{ij\alpha\beta}\Tilde{t}_{ij,\alpha\beta} c^\dagger_{i\alpha} c_{j\beta} + U\sum_i n_{i\uparrow} n_{i\downarrow},\label{eq:Hubbard model with SOI}
\end{equation}
where $\alpha, \beta = \uparrow,\downarrow$ represents the spin of the electrons, $i,j$ are site indices, and $n_{i\sigma} = c^\dagger_{i\sigma} c_{i\sigma}$ is the number operator for electrons on site $i$ with spin $\sigma$. The hopping matrix is spin-dependent in the presence of a SOC, and is given by $\Tilde{t}_{ij,\alpha\beta}$, where
\begin{equation}
\Tilde{t}_{ij,\alpha\beta} = \begin{cases} t_{ij}\delta_{\alpha\beta} + i\mathbf{v}_{ij}\cdot\bm{\sigma}_{\alpha\beta} &  i,j \, \text{nearest neighbours} \\ 0 & \text{otherwise.} \end{cases}\label{eq:Hopping with SOI}
\end{equation}
Here $\mathbf{v}_{ij}$, quantifies the effect of the spin-orbit interaction and $U$ is the on-site Coulomb repulsion. Additionally,
the hermiticity and time-reversal symmetry of Hamiltonian in Eq.~\eqref{eq:Hubbard model with SOI}, leads to a real symmetric $t_{ij}$ with $t_{ij}=t_{ji}$, and a pseudo-vector $\mathbf{v}_{ij}$, with $\mathbf{v}_{ij}= - \textbf{v}_{ji}$. The pseudo-vector $\mathbf{v}_{ij}$ breaks the inversion symmetry of Hamiltonian in Eq.~\eqref{eq:Hubbard model with SOI}. The broken inversion symmetry leads to a DM interaction in the effective spin model as in Sec. \ref{sec:model_eff}.

\subsection{Effective Spin Hamiltonian in the weak Mott insulating regime}\label{sec:model_eff}
\begin{figure}
    \centering
    \includegraphics{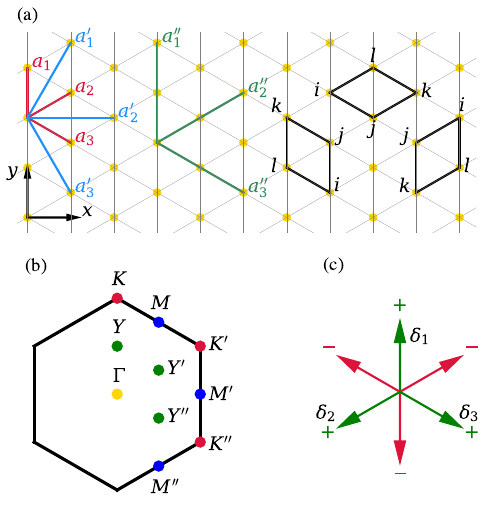}
    \caption{ (a) The triangular lattice spanned by the primitive vectors $\mathbf{a}_1=(0,1)$, $\mathbf{a}_2=(\sqrt{3}/2,1/2)$. The naming convention for the different types of bonds is shown. The three types of rhombuses (rings) are labeled by $(ijkl)$ counted anti-clockwise where $i$ is a sharp vertex of the rhombus.
    (b) Brillouin zone of the triangular lattice and some of the high-symmetry points. (c) Sign of the $\mathbf{v}_{ij}$ vector on moving along different directions in the triangular lattice, green is positive, red is negative. $\mathbf{\delta}_n$ are the three positive directions of $\mathbf{v}_{ij}$.}
    \label{fig:Fig-1}
\end{figure}

We are interested in the low-energy behaviour of the Hamiltonian in Eq.~\eqref{eq:Hubbard model with SOI} in the weak Mott insulating regime, and hence we perform a Schrieffer-Wolff canonical transformation \cite{SW_transformation} working in the regime $\Tilde{t}/U \ll 1$. The canonical transformation leads to a low-energy effective spin Hamiltonian by eliminating charge fluctuations to the high-energy doubly occupied sector. The effective Hamiltonian is of the form
\begin{equation}
H_\text{eff} = e^{iS}H e^{-iS},\label{eq:canonical_transformation}
\end{equation}
where $S$ is the generator of the canonical transformation. The expansion of Eq.~\eqref{eq:canonical_transformation} leads to a series of nested commutators, 
\begin{gather}
    H_{\text{eff}}=H+i\left[S,H \right]+\frac{i^2}{2}\left[S,\left[ S, H \right]\right]+\cdots,\label{eq:nest_comm}
\end{gather}
and $S$ is computed order by order in $\Tilde{t}/U$ where $S^{(n)} \propto (\Tilde{t}/U)^n$. At each order, $S^{(n)}$ is chosen such that $H_{\text{eff}}$ does not have any term which takes it out of the singly occupied sector. $S$ is computed up to $\mathcal{O}(\Tilde{t}^3/U^3)$, and the strong-coupling expansion in $\Tilde{t}/U$ in Eq.~\eqref{eq:nest_comm} is truncated to $\mathcal{O}(\Tilde{t}^4/U^3)$. (See Appendix \ref{sec:can-trans} for details of the canonical transformation). Finally, the resulting Hamiltonian is projected onto the singly-occupied subspace of the triangular lattice to obtain the effective spin model. For simplicity we have also considered $\mathbf{v}_{ij}=v_{z;i,j} \hat{\mathbf{z}}$ (out of plane). This choice is reflected in the symmetry of the resulting effective spin Hamiltonian in Eq.~\eqref{eq:eff_spin_model}, which has a $U(1)$ symmetry. Further, we have considered a uniform nearest neighbour hopping $t$, and a uniform $v_z$ with its positive direction on the triangular lattice as defined in Fig.~\ref{fig:Fig-1}(c). Along its positive (negative) direction $\mathbf{v}$ points along $\hat{\mathbf{z}}$ ($-\hat{\mathbf{z}}$). This choice of directions was motivated by studies on organic charge-transfer salts \cite{Yamashita_2011,Watanabe_2012,Minoru_2010,Zhao_2021} with the space group C2/c that have a DM interaction due to inversion symmetry breaking. 

In this study, we assume that $|\mathbf{v}|/t<1$ and we restrict ourselves to the leading order SOC-mediated terms in the Hamiltonian (see Appendix \ref{sec:full-model} for more details of the complete model). The resulting simplified spin model on a triangular lattice is 
\begin{gather}
    H_{\text{eff}}=\sum_{\langle ij \rangle}J_1 \mathbf{S}_i\cdot \mathbf{S}_j
     +\sum_{\langle\langle ij \rangle \rangle}J_2 \mathbf{S}_i\cdot \mathbf{S}_j
     +\sum_{\langle \langle\langle ij \rangle \rangle \rangle}J_3 \mathbf{S}_i\cdot \mathbf{S}_j \nonumber\\
     +\sum_{\langle ij \rangle} D_{z} {S}_i^{[x}{S}_j^{y]}+H_{\text{ring}}+H_{\text{ring,SOC}},\label{eq:eff_spin_model}
\end{gather}
where ${S}_i^{[x}{S}_j^{y]}=S_i^xS_j^y-S_i^yS_j^x$ is the anti-symmetrized product. In the DM interaction, only the nearest neighbour bonds oriented along the positive direction of $\mathbf{v}_{ij}$ are considered to avoid double-counting. The positive-oriented bonds are called $\mathbf{\delta}_n=(\mathbf{a}_1,-\mathbf{a}_2,\mathbf{a}_3)$ (see Fig.~\ref{fig:Fig-1} (a) \& (c)). The conventional ring-exchange term $H_{\text{ring}}$, is obtained from the $\Tilde{t}/U$ expansion at $\mathcal{O}(\Tilde{t}^4/U^3)$ by neglecting any effects of the SOC. It is expressed as:
\begin{gather}
     H_{\text{ring}}=\sum_{(ijkl) \in R}J_r  \left( \left( \mathbf{S}_i\cdot \mathbf{S}_j \right) \left( \mathbf{S}_k\cdot \mathbf{S}_l \right)
     +\left( \mathbf{S}_j\cdot \mathbf{S}_k \right) \left( \mathbf{S}_l\cdot \mathbf{S}_i \right)\right. \nonumber\\
     -\left. \left(\mathbf{S}_i\cdot \mathbf{S}_k \right) \left( \mathbf{S}_j\cdot \mathbf{S}_l \right) \right),
\end{gather}
where the sum is over the three types of rhombuses (rings) in the triangular lattice, where $i$ is a sharp vertex of the rhombus $(i,j,k,l)$ counted anti-clockwise (Fig.~\rev{1} (a)). The SOC generates two types of spin-bilinear terms \cite{Spin_triplet_VBS}, the DM interactions $D_{z} {S}_i^{[x}{S}_j^{y]}$, and anisotropic symmetric exchange interactions $S_i^a\Gamma^{ab}S^b_j$. Additionally, there are four types of SOC-mediated ring exchange terms, that are of $\mathcal{O}\left(t^4/U^3\left(v_z/t \right)^n\right)$, where $n \in \{1,2,3,4$\} (see Appendix \ref{sec:full-model}). The leading order SOC-mediated ring-exchange term at $\mathcal{O}\left(t^4/U^3 \left(v_z/t \right) \right)$ is:
\begin{gather}    
H_{\text{ring,SOC}}=\sum_{(ijkl) \in R}J_{r_1}\left( 2{S}_j^{[x}{S}_l^{y]}{S}_i^{z} {S}_k^{z}\right.\nonumber\\
    \left.+{S}_i^{[x}{S}_j^{y]} \left({S}_k^x{S}_l^x+{S}_k^y{S}_l^y\right)+{S}_l^{[x}{S}_i^{y]}\left({S}_j^x{S}_k^x+{S}_j^y{S}_k^y\right)\right.\nonumber\\
    \left.-{S}_j^{[x}{S}_k^{y]} \left({S}_l^x{S}_i^x+{S}_l^y{S}_i^y\right) 
    -{S}_k^{[x}{S}_l^{y]} \left({S}_i^x{S}_j^x+{S}_i^y{S}_j^y\right) \right).\label{eq:H_SOI_mediated_ring_1}
\end{gather}
The higher-order ($n>1$) SOC-mediated ring-exchange terms and further neighbour DM terms have been neglected as we are working in the regime of a weak SOC ($v_z/t<1$). We also do not consider the $\mathcal{O}\left((v_z/t)^2 \right)$ corrections to the Heisenberg exchanges. Additionally, we have neglected all the n-th neighbour anisotropic symmetric exchange interactions $\Gamma_n$. This is justified in a regime where $\Gamma_n<J_{r_1}$ (see Appendix \ref{sec:full-model}). Hence, in our simplified spin model the exchange interactions are parameterized in terms of $t/U$ and $v_z/t$.
\begin{gather}
    J_1=\frac{4t^2}{U}-\frac{28t^4}{U^3}, \quad J_2=\frac{4t^4}{U^3}, \quad J_3=\frac{4t^4}{U^3},\nonumber \\
    J_r=\frac{80t^4}{U^3},\quad D_z=\frac{8t v_z}{U}, \quad J_{r_1}=\frac{160 t^3 v_z}{U^3}.\label{eq:Exchanges_from_SC_exp}
\end{gather}
Later while constructing the quantum phase diagrams in Sec. \ref{sec:QPD} we treat $D_z$ and $J_{r_1}$ as independent parameters.

\section{Quantum Phase Diagrams}\label{sec:QPD}

In Sec. \ref{sec:model_eff}, we discussed that the dominant contribution of the SOC to the effective spin Hamiltonian in Eq.~\eqref{eq:eff_spin_model} are from the nearest-neighbour DM term and a SOC-mediated ring-exchange interaction (that is leading order in $v_z/t$). We expect the DM interaction to favour a magnetically ordered $120^{\circ}$ ground state \cite{Hog_2022}. However, the SOC-mediated ring-exchange term, like the conventional ring-exchange, can potentially lead to increased frustration that could result in exotic quantum ground states, like a CSL. In this section, we investigate whether compensation between these two kinds of terms can help to stabilize a quantum phase. Henceforth, we treat $D_z$ and $J_{r_1}$ as independent parameters to isolate their effects. We use two different values of $t/U$ so that in the absence of a SOC the systems are in the CSL and VBS phases respectively. We then investigate the phases that appear as the strengths of $D_z$ and $J_{r_1}$ are tuned.

For constructing the quantum phase diagrams we have used the infinite-DMRG (iDMRG) algorithm \cite{White_1992,White_1993,RevModPhys_Schollwock,mcculloch_2008,Vidal_2008} to calculate the quantum ground states. iDMRG was implemented using the Python package TeNPy \cite{Hauschild_TeNPy_2018}. iDMRG variationally optimizes a 1D matrix product state (MPS) representation of the wavefunction, to converge to the correct ground state. In our simulations, we have used a cylindrical geometry of finite circumference and infinite length. For the cylinder, the YC orientation was chosen where one of the sides is parallel to the circumference. We used a circumference length of $L_y=6$ in the $\mathbf{a}_1$ direction (see Fig.~\ref{fig:Fig-1} (a)). An MPS unit cell of length $L_x=2$ in the infinite $\mathbf{a}_2$ direction was used, as a third nearest neighbour interaction on a triangular lattice could be accommodated within this cell. iDMRG enables us to calculate long-range correlations along this infinite direction (axis of the cylinder). The spin model in Eq.~\eqref{eq:eff_spin_model} has a $U(1)$ symmetry, and this symmetry is built into the MPS representation. Correspondingly, $S^z_{\text{tot}}=\sum_i S^z_{i}$ is conserved, as $\left[ S^z_{\text{tot}}, H_{\text{eff}} \right]=0$. In this study, we leverage this symmetry numerically, and focus only on the $S^z_{\text{tot}}=0$ sector. In all our simulations we have used a bond dimension $\chi_{\text{max}}=1600$. (See Appendix \ref{sec:DMRG_cals} for details of the iDMRG implementation.)

\begin{figure*}
    \centering
    \hspace{-0.6cm}
\includegraphics{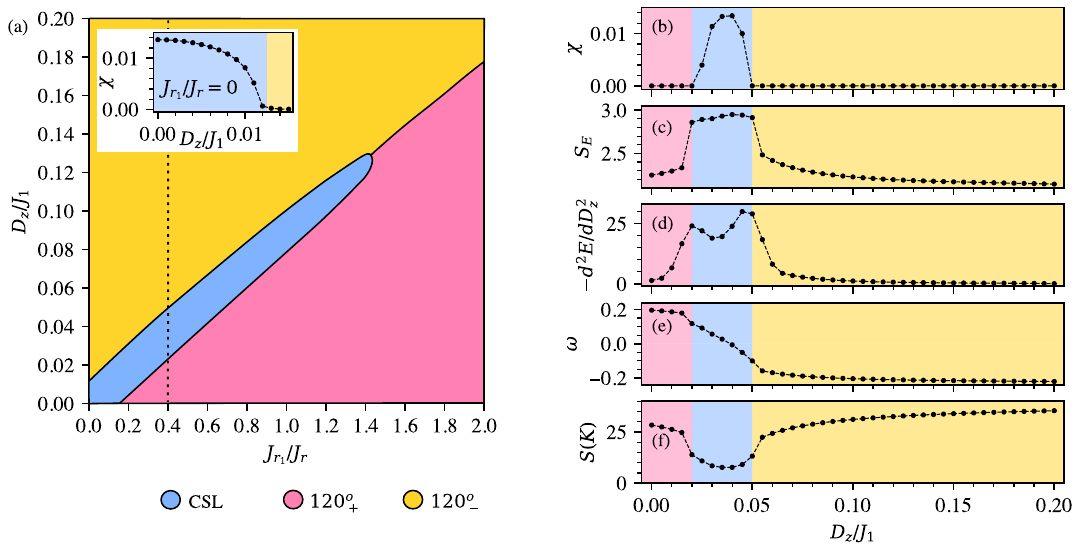}
    \caption{(a) Quantum Phase diagram obtained from iDMRG by starting from a CSL phase at $J_1=1$, $J_2=J_3=0.01$, $J_r=0.2$ ($t/U=0.097$). The SOC generates the nearest neighbour DM interaction $D_z$ and a SOC-mediated ring exchange interaction $J_{r_1}$. The two types of $120^{\circ}$ orders in (a) are distinguished by their handedness $\omega$. The $D_z$ term prefers the  $120^{\circ}_{-}$ phase with $\omega<0$ and the $J_{r_1}$ term prefers $120^{\circ}_{+}$ phase with $\omega>0$. The compensation between these opposing tendencies leads to the stability of the CSL phase in a narrow diagonal region, where their combined effect is minimized. Along the vertical axis $J_{r_1}/J_r=0$, the CSL phase persists until $D_z/J_1=0.0125$, as in inset. Observables along the vertical dashed line at $J_{r_1}/J_r=0.4$ are shown in (b)-(f).  The CSL has a nonzero scalar chirality $\chi$, a high entanglement entropy $S_E$, and a weak intensity at $S(K)$, compared to the $120^{\circ}_{\pm}$ phases. Location of the peaks in the second derivative of the energy in (d), can also be used to identify the location of the phase boundaries.
    }
    \label{fig:Fig-2_CSL_phase_diagram}
\end{figure*}

\subsection{Starting in chiral spin liquid phase ($t/U=0.097$)}\label{sec:Start_CSL}
A CSL is a type of QSL that spontaneously breaks time-reversal and parity symmetries \cite{Laughlin_1987, Laughlin_1990}. At the mean-field level the CSL can be described using a fermionic parton construction \cite{Wen_2002,Savary_2016,RevModPhys_Zhou}, where the electrons fractionalize into neutral spin $1/2$ particles. These fermionic partons are in bands with non-zero Chern numbers and are coupled to an emergent $U(1)$ gauge field. The $U(1)$ gauge field is gapped by a Chern-Simons term, and this leads to non-trivial mutual statistics \cite{Laughlin_1989,Laughlin_1990} of the deconfined spinon excitations in the CSL. Additionally, the CSL carries propagating chiral edge modes \cite{Wen_1991}, similar to quantum Hall systems \cite{Laughlin_1987,Laughlin_1990}.

Due to spontaneous time-reversal symmetry breaking in the CSL state, the scalar chirality acquires a non-zero expectation value \cite{Wen_1989}. The scalar chirality averaged over all triangles is defined as
\begin{gather}
    \chi=\frac{1}{N_t}\sum_{i,j,k \in \lhd, \rhd} \left\langle \mathbf{S}_i\cdot \left(\mathbf{S}_j \times \mathbf{S}_k  \right) \right\rangle,\label{eq:chi}
\end{gather}
where $i,j,k$ are taken anti-clockwise, and $N_t=2 L_x L_y$ is the total number of triangles in the MPS unit cell. However, a non-zero scalar chirality can also indicate a non-coplanar order. More conclusive evidence of the CSL phase is found from the momentum-resolved entanglement spectrum which breaks inversion/time-reversal symmetry. Additionally the Kalmeyer-Laughlin type CSL, has a characteristic degeneracy pattern in the entanglement spectrum: $\{1,1,2,3,\cdots\}$ with increasing momenta, for $|S_z|=\{0,1,2,3,\cdots\}$ \cite{Wen_1991,Li_2008}.  In our study, a non-zero scalar chirality, correct degeneracy structure of the entanglement spectrum, and a lack of long-range magnetic order have been used to identify the CSL phase.

We first establish the nature of the ground state in the absence of a SOC. We use $t/U=0.097$, which from Eq.~\eqref{eq:Exchanges_from_SC_exp} corresponds to $J_1=1$, $J_2=0.01$, $J_3=0.01$, $J_r=0.2$. The ground state obtained using iDMRG using this parameter set is in the CSL phase. This result is in agreement with \cite{schultz2023electric}. 

Next, we characterize the ground state in the presence of a SOC. We observe in Fig.~\ref{fig:Fig-2_CSL_phase_diagram} (a), that with increasing the strengths of the SOC, $D_z$ and $J_{r_1}$, the CSL phase survives in a narrow diagonal region, until $D_z/J_1=0.13, J_{r_1}/J_r=1.4$. Therefore the CSL phase is stabilized through a combined effect of $D_z$ and $J_{r_1}$ in the presence of a SOC. We also identify two different types of $120^{\circ}$ phases, above and below the CSL phase in Fig.~\ref{fig:Fig-2_CSL_phase_diagram} (a). The $120^{\circ}$ phases have long-range magnetic order and can be identified from the sharp peaks in the static spin structure factor
\begin{gather}
    S(\mathbf{k})=\sum_{ij}\left(\left\langle \mathbf{S}_i\cdot \mathbf{S}_j \right\rangle -\left\langle \mathbf{S}_i \right\rangle\cdot \left\langle \mathbf{S}_j \right\rangle \right)e^{i\mathbf{k}\cdot \left( \mathbf{R}_i-\mathbf{R}_j\right)},\label{eq:STF}
\end{gather}
at the $K$ points (see Figs.~\ref{fig:Fig-1} (b), \ref{fig:120p} \& \ref{fig:120m}) in the Brillouin zone. Classically, the $120^{\circ}$ order has a $3$-site unit cell, and the spins are co-planar having a relative angle of $120^{\circ}$ between each other. Unlike the classical phase, the quantum $120^{\circ}$ phase is highly entangled, but has similar long-ranged correlations as the classical $120^{\circ}$. To classify the two different types of $120^{\circ}$ phases in Fig.~\ref{fig:Fig-2_CSL_phase_diagram}, we further subdivide the quantum $120^{\circ}$ phase into the $120^{\circ}_{+}$ and $120^{\circ}_{-}$ phase based on their handedness. The handedness $\omega$ is defined as:
\begin{gather}
    \omega=\frac{1}{N_b}\sum_{i,\mathbf{\delta}_n} \left\langle{S}^{[x}_i {S}^{y]}_{i+\mathbf{\delta}_n}\right\rangle,\label{eq:handedness}
\end{gather}
where $N_b=3L_xL_y$ is the number of nearest neighbour bonds in the triangular lattice. We define the $120^{\circ}_{+}$ phase to have $\omega>0$, and $120^{\circ}_{-}$ to have $\omega<0$, in addition to sharp peaks at $S(K)$ and long-range spin correlations. The $120^{\circ}$ ordered phase appearing in the absence of a SOC has $\omega=0$, this is the type of $120^{\circ}$ order observed in iDMRG studies \cite{Cookmeyer_four_spin,schultz2023electric}. More details regarding the different types of $120^{\circ}$ ordered phases are discussed in Appendix \ref{sec:120pm_details}.

We can understand the stability of the CSL phase as a result of compensation between two opposing tendencies. Above the diagonal CSL (blue) region in Fig.~\ref{fig:Fig-2_CSL_phase_diagram} (a), the ordering tendency of the DM interaction overcomes the frustration induced by the other terms. For large values of $D_z$, an $\omega<0$ or $120^{\circ}_{-}$ phase is preferred as it minimizes the energy; the definition of $\omega$ is similar to the Hamiltonian of the DM interaction. Below the diagonal CSL region in Fig.~\ref{fig:Fig-2_CSL_phase_diagram}  (a), the SOC-mediated ring-exchange interaction is dominant. However, unlike the conventional ring exchange, the SOC-mediated ring exchange interaction prefers a magnetically ordered $120^{\circ}_{+}$ phase with $\omega>0$. This can be understood using a heuristic argument presented in Sec. \ref{sec:Stability_CSL}. Hence, there is a competition between the DM interaction $D_z$ and SOC-mediated ring exchange interaction $J_{r_1}$, each of which prefer a differently handed magnetically ordered $120^{\circ}$ phase. It is as a result of the compensation between these two interactions that the CSL is stabilized in the elongated diagonal region where the two opposing tendencies effectively cancel out. Therefore we find that the CSL phase is robust to small perturbations due to a SOC, as is apparent from its stability along the diagonal region in Fig.~\ref{fig:Fig-2_CSL_phase_diagram} (a).

\begin{figure*}
    \centering
    \hspace{-0.6cm}
    \includegraphics{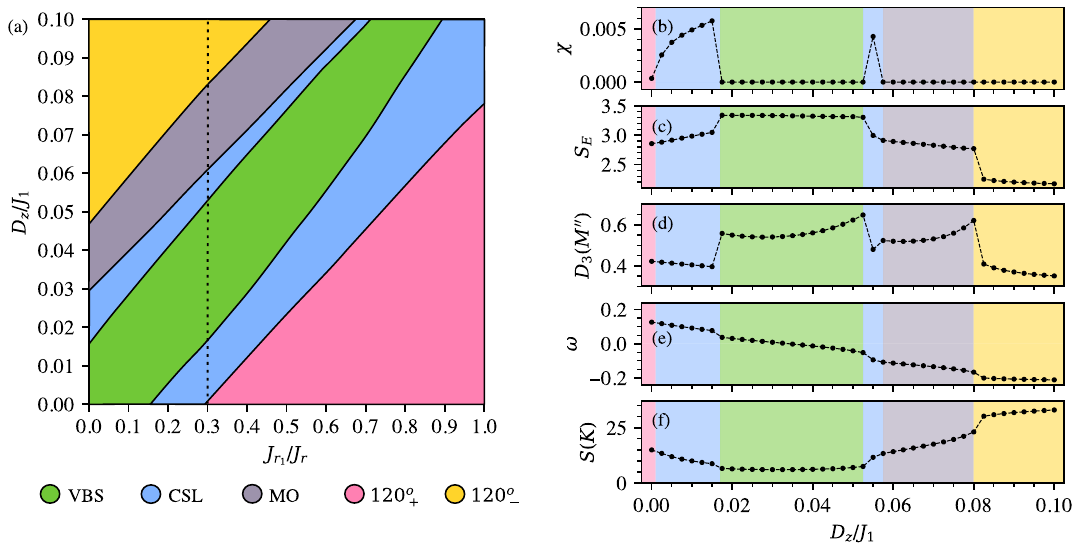}
    \caption{(a) Quantum Phase diagram obtained from iDMRG by starting from a VBS phase at $J_1=1$, $J_2=J_3=0.012$,$J_r=0.24$ ($t/U=0.105$).  
    Similar to Fig.~\ref{fig:Fig-2_CSL_phase_diagram}, the compensation between the DM interaction $D_z$ and SOC mediated-ring exchange $J_{r_1}$ leads to the VBS phase being stabilized in the diagonal region. In addition to the $120^{\circ}_{\pm}$ orders, there are narrow regions of the CSL accompanying the VBS. Additionally, a magnetic ordered (MO) phase with long-range correlations, distinct from the $120^{\circ}_{-}$ is identified. Observables along the vertical dashed line $J_{r_1}/J_r=0.3$ are shown in (b)-(f). The VBS is characterized by a weak intensity at $S(K)$ and strong peaks in the dimer structure factor $D_3(M'')$. To clearly distinguish the CSL around $D_z/J_1 \sim 0.055$ from its neighbouring states, a zoomed-in plot is provided in Fig.~\ref{fig:zoom}.}
    \label{fig:Fig-3_VBS_phase_diagram}
\end{figure*}

\subsection{Starting in valence bond solid phase ($t/U=0.105$)}
A VBS is formed from a covering of singlets of nearest neighbour spins \cite{Anderson_1973,Anderson_1987,Anderson_1987_2,Affleck_1987} with all the singlet bonds occurring along a particular direction in the lattice. The VBS can be identified from sharp peaks in the static dimer structure factor
\begin{gather}
    D_n(\mathbf{k})=\sum_{ij}\left(  \left\langle D^n_iD^n_j \right\rangle-\left\langle D^n_i \right\rangle \left\langle D^n_j \right\rangle \right) e^{i\mathbf{k}\cdot \left( \mathbf{R}_i-\mathbf{R}_j\right)},\label{eq:DSF}
\end{gather}
where $D^n_i=\mathbf{S}_i\cdot \mathbf{S}_{i+\mathbf{a}_n}$ is the dimer operator along the $\mathbf{a}_n$ direction (see Fig.~\ref{fig:Fig-1} (a)). The peak in the static dimer structure factor occurs at the $M$ point corresponding to the translation vectors of the singlet covering bonds. However, a VBS state obtained from iDMRG is peaked at $M$ points in the dimer structure factor along all three directions $\mathbf{a}_n$. This is because the ground state obtained from iDMRG is a superposition of different types of singlet dimer coverings. The dominant dimer covering can be identified from the positions of the peaks and their relative intensities \cite{schultz2023electric}.

\begin{figure}
    \centering
    \includegraphics[scale=1.2]{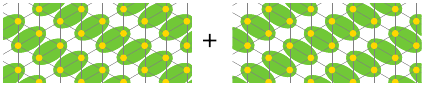}
    \caption{The two equal-weight dominant dimer coverings in the VBS phase. The singlet bonding axes in the dimer coverings are along $\mathbf{a}_2$ and $\mathbf{a}_3$ respectively.}
    \label{fig:VBS_state}
\end{figure}

First, we characterize the nature of the ground state in the absence of a SOC when $t/U=0.105$.
From Eq.~\eqref{eq:Exchanges_from_SC_exp}, this value corresponds to $J_1=1$, $J_2=0.01$, $J_3=0.01$, $J_r=0.24$. With this parameterization, we find a VBS ground state, in agreement with \cite{schultz2023electric}. The peaks of the dimer structure factor at $D_2(M)$ and $D_3(M')$ have the highest intensity (see Figs.~\ref{fig:Fig-1} (a) \& \ref{fig:VBS}). This implies the VBS state is, to leading order, a superposition of two types of dimer coverings \cite{schultz2023electric}. These two dimer coverings have their singlet bonding axes oriented along $\mathbf{a}_2$ and $\mathbf{a}_3$ respectively, in the triangular lattice (see Fig. \ref{fig:VBS_state}).

In the presence of a SOC, we find from Fig.~\ref{fig:Fig-3_VBS_phase_diagram} (a) that the VBS phase is stabilized along the diagonal region as the strengths of $D_z$ and $J_{r_1}$ are increased. This stabilization can be attributed to the same mechanism which is a result of compensation between the DM interaction and SOC-mediated ring exchange interactions as discussed in Sec. \ref{sec:Start_CSL}. Interestingly in Fig.~\ref{fig:Fig-3_VBS_phase_diagram} (a), we also observe two bands of the CSL phase, above and below the VBS phase. These CSL bands are stabilized along a diagonal region by a similar mechanism. However, states in these two CSL bands have opposite signed handedness $\omega$. We find that increasing the strength of $J_{r_1}$ drives a VBS state into a CSL, we provide a qualitative explanation of this fact in Sec. \ref{sec:Stability_VBS}. Apart from observing the long-range ordered $120^{\circ}_{-}$ and $120^{\circ}_{+}$ phases in similar regimes compared to Fig.~\ref{fig:Fig-2_CSL_phase_diagram} (a), we also observe another magnetically ordered (MO) phase in Fig.~\ref{fig:Fig-3_VBS_phase_diagram} (a). Even though the MO phase is long-range ordered, it has a smaller correlation length than the $120^{\circ}_{-}$ (see Figs.~\ref{fig:120m} \& \ref{fig:MO}). Additionally, the MO phase has a higher entanglement entropy $S_E$, and a higher intensity at $D_3(M'')$ compared to the the $120^{\circ}_{-}$ phase. We are unable to fully characterize the MO phase based on the available data from the iDMRG simulations. The MO phase may be related to the AFM-II phase identified in a variational Monte Carlo study \cite{Zhao_2021} of the $J_1$-$J_r$-$D_z$ model. The AFM-II phase in the study \cite{Zhao_2021} is predicted to have peaks in the static spin structure factor close to but not exactly at the $K$ point. However, we could not resolve a similar feature for the MO phase from our iDMRG simulations.


\section{Insights into the phase diagrams from qualitative arguments}\label{sec:mft}

In this section, we present some qualitative explanations for the stability of the CSL and VBS phases shown in the quantum phase diagrams in Figs.~\ref{fig:Fig-2_CSL_phase_diagram} \& \ref{fig:Fig-3_VBS_phase_diagram}. These arguments provide some intuition behind the stabilization mechanism and the fact that there is an extended region of stability for these phases.


\subsection{Stability of the chiral spin liquid and valence bond solid phases}\label{sec:Stability_CSL}
We observe from Figs.~\ref{fig:Fig-2_CSL_phase_diagram} \& \ref{fig:Fig-3_VBS_phase_diagram}, that, upon simultaneously increasing the strength of $D_z$ and $J_{r_1}$, the CSL and VBS phases are stabilized in respective elongated diagonal regions. Such an elongated diagonal region of stability is indicative of compensation between the SOC-mediated ring-exchange and the DM interaction at a particular ratio of $D_z/J_{r_1}$. To investigate this, we look at an effective Hamiltonian in two different limits:  above and below the compensation regions, which are the blue region (CSL) in Fig.~\ref{fig:Fig-2_CSL_phase_diagram} (a), or green region (VBS) in Fig.~\ref{fig:Fig-3_VBS_phase_diagram} (a).

\begin{itemize}
\item Case A: In the limit of $D_z/J_{r_1}\gg 1$, we ignore the effect of the SOC-mediated ring exchange $J_{r_1}$, and therefore the $H_{\text{eff,A}}=H_{D_z}=\sum_{\langle i,j \rangle } D_z {S}_i^{[x}{S}_j^{y]}$.
\item Case B: In the opposite limit of $J_{r_1}/D_z\gg 1$, we ignore the $D_z$ term, and $H_{\text{eff,B}}=H_{J_{r_1}}$ from \eqref{eq:H_SOI_mediated_ring_1}. To write $H_{J_{r_1}}$ in a form resembling the DM interaction, we assume co-planar classical spins (with $S=1/2$) in the $120^{\circ}_{+}$ order ($120^{\circ}$ angle between neighbouring spins). Then, we can express $H_{\text{eff,B}}\approx -\frac{1}{2}\sum_{\langle i,j \rangle} J_{r_1} {S}_i^{[x}{S}_j^{y]}$. Here we have used a heuristic decoupling of the four-spin SOC-mediated ring exchange term as
\begin{gather}
    {S}_i^{[x}{S}_j^{y]}\left( S^x_k S^x_l+S^y_k S^y_l \right)\rightarrow {S}_i^{[x}{S}_j^{y]} \left\langle S^x_k S^x_l+S^y_k S^y_l\right\rangle, \label{eq:H_MF_CSL_step}
\end{gather}
and since the spins are coplanar we have used $\left\langle S^x_k S^x_l+S^y_k S^y_l\right\rangle=\mathbf{S}_k\cdot \mathbf{S}_l=|S|^2\cos(2\pi /3)=-1/8$. The overall negative sign in $H_{\text{eff,B}}$ is consistent with the stability of the $120^{\circ}_{+}$ phase in this regime. This decoupling occurs for 4 different rhombuses having the bond $(i,j)$, leading to the factor of $-1/2$ in $H_{\text{eff,B}}$ (see Fig.~\ref{fig:MFT-lattice} (a)).
\end{itemize}

\begin{figure}
    \centering
    \hspace{-0.5cm}
    \includegraphics{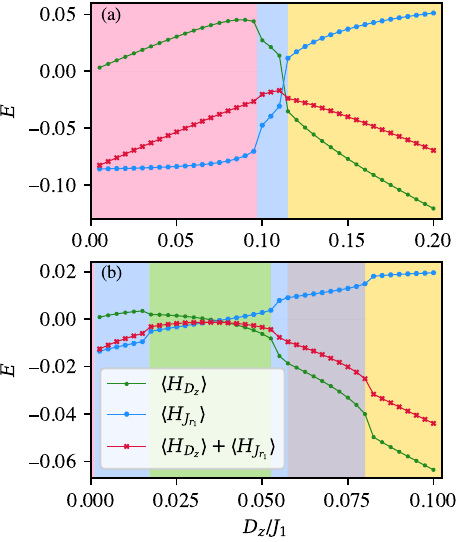}
    \caption{The expectation values of $\langle H_{D_z}\rangle $, $\langle H_{J_{r_1}}\rangle $ and $\langle H_{D_z}+ H_{J_{r_1}}\rangle $ in the ground state calculated along: (a) $J_{r_1}/J_r=1.2$ in Fig.~\ref{fig:Fig-2_CSL_phase_diagram}, and (b) $J_{r_1}/J_r=0.3$ in Fig.~\ref{fig:Fig-3_VBS_phase_diagram}. The background colors denote the various phases in the corresponding figures. $\langle H_{D_z}+ H_{J_{r_1}}\rangle$ is close to zero (and its absolute value is minimized) in the region where the CSL (as in (a)) and the VBS (as in (b)) are stabilized, indicating that in this region $H_{D_z}$ and $H_{J_{r_1}}$ compensate for each other.}
    \label{fig:H_exp}
\end{figure}

Now that both limits are written in the form of a DM interaction, it is easy to identify a compensation between $H_{\text{eff,A}}$ and $H_{\text{eff,B}}$ when $H_{\text{eff,A}}+H_{\text{eff,B}} \approx 0$, that is $D_z/J_{r_1}=0.5$. Comparing with the iDMRG results from Figs.~\ref{fig:Fig-2_CSL_phase_diagram} \& \ref{fig:Fig-3_VBS_phase_diagram}, we find that in the region of stability $D_z/J_{r_1}\approx 0.44$ (for the CSL) and $D_z/J_{r_1}\approx 0.52$ (for the VBS) respectively, and are within $12 \%$ of that expected from the qualitative argument. Additional evidence in support of this stabilization mechanism can be found by examining the expectation value $\left\langle H_{D_z}+H_{J_{r_1}}\right\rangle$ at various locations in the phase diagram. In Figs.~\ref{fig:H_exp} (a) \& (b), we observe the absolute value of this quantity to be minimized and close to zero in the region of stability of the CSL and VBS phases, respectively. This observation further demonstrates the two SOC-mediated terms, $H_{D_z}$ and $H_{J_{r_1}}$, compensate for each other in the region of stability of these phases.

\subsection{Extent of the chiral spin liquid phase}\label{sec:Stability_CSL_b}

\begin{figure}
    \centering
    \includegraphics{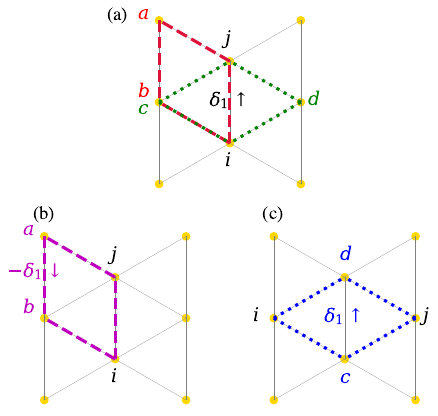}
    \caption{(a) One of the four rhombuses which has an edge $\langle a,b\rangle$ parallel to the bond $\langle i,j \rangle_{\mathbf{\delta}_1}$, oriented along $\mathbf{\delta}_1$, is shown in red. The rhombus in green has second nearest neighbour sites $(c,d)$ and has the bond $\langle i,j \rangle_{\mathbf{\delta}_1}$ as its diagonal. (b) One of the four rhombuses with the bond $\langle a,b\rangle_{-\mathbf{\delta}_1}$ oriented along $-\mathbf{\delta}_1$ that also has the bond $\langle i,j \rangle$ as a parallel edge. (c) A rhombus with the bond $\langle c,d\rangle_{\mathbf{\delta}_1}$ as its diagonal which has the second-nearest neighbour sites $(c,d)$.}
    \label{fig:MFT-lattice}
\end{figure}

In Fig.~\ref{fig:Fig-2_CSL_phase_diagram} (a), we find that the CSL phase is stabilized until $D_z/J_1=0.13, J_{r_1}/J_r=1.4$, even though $D_z/J_{r_1}\sim 0.5$ (the compensation condition discussed in Sec. \ref{sec:Stability_CSL}). This indicates a failure of the compensation mechanism beyond a certain value of $J_{r_1}$. To understand the reason for the failure, we construct a heuristic Hamiltonian for the CSL phase, using the DM and the SOC-mediated ring exchange interactions. We decouple the $H_{J_{r_1}}$ term the same way as in Eq.~\eqref{eq:H_MF_CSL_step}, where now the expectation value $\langle \cdot \rangle$ is computed with respect to an iDMRG wavefunction from CSL phase. After the decoupling we can construct an effective DM type interaction, containing both the original DM term as well as the correction from $H_{J_{r_1}}$. Conceptually, we are mapping the two independent variables $D_z$ and $J_{r_1}$ of the phase diagram in Fig.~\ref{fig:Fig-2_CSL_phase_diagram} (a) into one variable $D_z$. This makes sense for the CSL phase since it appears in a narrow elongated region (with $D_z/J_{r_1}\sim 0.5$) which can be approximated as being one-dimensional for the purpose of this discussion. The effective Hamiltonian is given by:

\begin{gather}
    H_{\text{eff}}=\sum_{\langle i,j \rangle}D_z S_i^{[x}S_j^{y]}\left( 1+ \frac{J_{r_1}}{D_z} \sum_{\langle a,b \rangle} \left\langle S^x_a S^x_b+S^y_a S^y_b\right\rangle \right.\nonumber\\
    \left.+\frac{2 J_{r_1}}{D_z}\sum_{\langle \langle c,d\rangle \rangle} \left\langle S^z_c S^z_d\right\rangle \right)+H_{\text{Heisenberg}}+H_{\text{ring}}.
\end{gather}
We quantify the correction to $D_z$ by defining $\Lambda_{\mathbf{\delta}_n}$ for the three types of bonds $\mathbf{\delta}_n$ on a triangular lattice (see Fig.~\ref{fig:Fig-1} (a)):
\begin{gather}
    \Lambda_{\mathbf{\delta}_n}=\frac{1}{L_x L_y} \sum_{\langle i,j \rangle_{\mathbf{\delta}_n}} \left( 1+\frac{J_{r_1}}{D_z} \sum_{\langle a,b \rangle} \left\langle S^x_a S^x_b+S^y_a S^y_b\right\rangle  \right.\nonumber\\
    \left.+\frac{2 J_{r_1}}{D_z}\sum_{\langle \langle c,d\rangle \rangle} \left\langle S^z_c S^z_d\right\rangle \right),
\end{gather}
where $\langle i,j \rangle_{\mathbf{\delta}_n}$ denotes nearest neighbour bonds oriented in the $\mathbf{\delta}_n$ direction (see Fig.~\ref{fig:Fig-1} (c) for conventions). We average over all bonds of a particular orientation within the MPS unit cell. $(a,b)$ are sites such that $( i,j,a,b )$ forms a rhombus (there are 4 such choices for $(a,b)$), and $(c,d)$ are sites such that $(i,d,j,c)$ forms a rhombus. An example is shown in Fig.~\ref{fig:MFT-lattice} (a), where the $\langle i,j \rangle_{\mathbf{\delta}_1}$ bond is shown, one of the four rhombuses having an edge $\langle a,b\rangle$ parallel to bond $\langle i,j \rangle_{\mathbf{\delta}_1}$ is shown in red, and the rhombus having second nearest neighbour sites $(c,d)$ such that the bond $\langle i,j \rangle_{\mathbf{\delta}_1}$ is its diagonal is shown in green.  We then average $\Lambda_{\mathbf{\delta}_n}$ over all of the bond directions $\mathbf{\delta}_n$.
\begin{gather}
    \Lambda_p=\frac{1}{3}\left( \Lambda_{\mathbf{\delta}_1}+\Lambda_{\mathbf{\delta}_2} +\Lambda_{\mathbf{\delta}_3}\right).
\end{gather}
Finally, we average $\Lambda_p$ over all CSL wavefunctions sampled by the underlying grid in our iDMRG simulations, $\Lambda=\langle \Lambda_p \rangle_{p \in \text{CSL}}$. We find in the heuristic Hamiltonian that the DM term is renormalized by $\Lambda$ within the CSL phase. Note that, strictly speaking, the renormalization is different at different points within the CSL phase; $\Lambda$ as we have defined it quantifies the \textit{average} renormalization of the DM interaction due to $J_{r_1}$ within the CSL phase. This leads us to the following effective Hamiltonian:
\begin{gather}
    H_{\text{eff, CSL}}=\sum_{\langle i,j \rangle} \Lambda D_z {S}_i^{[x}{S}_j^{y]} +H_{\text{Heisenberg}}+H_{\text{ring}}.\label{eq:H_MF_CSL}
\end{gather}
If we had $\Lambda=0$, then this would indicate perfect compensation. However, we find $\Lambda=0.1284$. This nonzero value of $\Lambda$ accounts for the eventual disappearance of the CSL phase, even when $D_z/J_{r_1}\sim 0.5$. From inset in Fig.~\ref{fig:Fig-2_CSL_phase_diagram} (a), we know from exact iDMRG simulations that in the absence of SOC-mediated ring-exchange ($J_{r_1}=0$), the DM interaction $D_z$ destabilizes the CSL phase beyond $D_z/J_1= 0.0125$. For a renormalized DM interaction $\Lambda D_z$, as in our heuristic Hamiltonian in Eq.~\eqref{eq:H_MF_CSL}, this destabilization would happen beyond $D_z/J_1= 0.0125/\Lambda \approx 0.097$. Hence, our heuristic Hamiltonian for the CSL phase predicts that the CSL phase should be stabilized until $D_z/J_1 \approx 0.097$. From iDMRG of the full model in Eq.~\eqref{eq:eff_spin_model}, we found that the  CSL phase persists until $D_z/J_1=0.13$. The value predicted from the heuristic model is within $26 \%$ of the exact value obtained from the iDMRG simulations. Therefore, the heuristic model can qualitatively describe the extent of the CSL phase in Fig.~\ref{fig:Fig-2_CSL_phase_diagram} (a).

\begin{figure}
    \centering
    \hspace{-0.5cm}
    \includegraphics{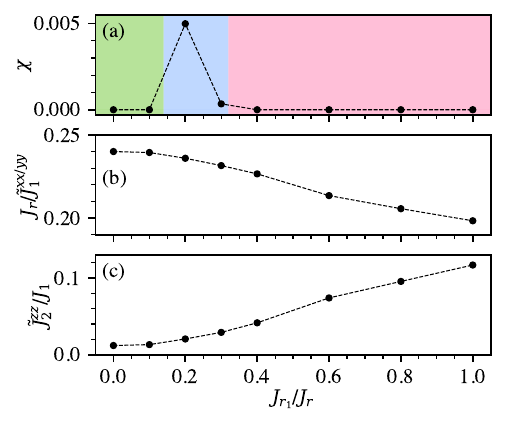}
    \caption{(a) The scalar chirality $\chi$ along the axis $D_z=0$ of Fig.~\ref{fig:Fig-3_VBS_phase_diagram} a. The CSL phase is centered around $J_{r_1}/J_r\approx 0.2$. The background colours denote the corresponding phases in Fig.~\ref{fig:Fig-3_VBS_phase_diagram}. With increasing $J_{r_1}/J_r$ the heuristic model in Sec. \ref{sec:Stability_VBS} predicts that $J_r/\Tilde{J}^{xx/yy}_1$ decreases (as in (b)) and $\Tilde{J}_2^{zz}/J_1$ increases (as in (c)). Starting from a VBS phase, these trends indicate that a CSL phase should appear followed by a $120^{\circ}$ ordered phase with increasing values of $J_{r_1}/J_r$.}
    \label{fig:MFT-VBS}
\end{figure}

\subsection{Inducing a chiral spin liquid starting from a valence bond solid phase}\label{sec:Stability_VBS}

In Fig.~\ref{fig:Fig-3_VBS_phase_diagram} (a), we observe two narrow bands of CSL (in blue) adjacent to the VBS phase. In this section, we give a qualitative explanation for the appearance of the CSL phase upon increasing $J_{r_1}$, when starting from the VBS phase. In particular, we construct a heuristic Hamiltonian for states along the $D_z=0$ axis. Starting from a VBS at $J_{r_1}/J_r=0$ ,
the CSL appears with increasing $J_{r_1}$ when $J_{r_1}/J_r\approx 0.2$ as in Fig.~\ref{fig:Fig-3_VBS_phase_diagram}  (a). We know from the literature \cite{Cookmeyer_four_spin,schultz2023electric} that, when starting from the VBS phase, decreasing $J_r/J_1$ and increasing $J_2/J_1$ first favours the CSL phase, followed by the $120^{\circ}$ order. We want to see if we can capture this trend using a heuristic model with renormalized Heisenberg couplings.
Along the $D_z=0$ axis, the only SOC-mediated term in the Hamiltonian is the four-spin SOC-mediated ring-exchange $H_{J_{r_1}}$. We decouple this four-spin term as 
\begin{gather}
    \hspace{-0.18 cm}\left( S^x_i S^x_j+S^y_i S^y_j \right){S}_k^{[x}{S}_l^{y]}\rightarrow  \left( S^x_i S^x_j+S^y_i S^y_j\right) \langle {S}_k^{[x}{S}_l^{y]} \rangle,\label{eq:H_MF_VBS_step}
\end{gather}
where the expectation value $\langle \cdot \rangle$ is computed in the ground state using the iDMRG wavefunctions at these particular parameters (this is different from Sec. \ref{sec:Stability_CSL_b}, where an average over different points in the phase diagram was taken). Next, we compute $\Delta_{1,\mathbf{\delta}_n}$ and $\Delta_{2,\mathbf{\delta}_n}$ for each of the three types of bonds $\mathbf{\delta}_n$ in the triangular lattice:
\begin{gather}
\Delta_{1,\mathbf{\delta}_n}=\frac{1}{L_x L_y}\sum_{\langle i,j \rangle}\sum_{\langle a,b\rangle_{-\mathbf{\delta}_n}} \langle {S}_a^{[x}{S}_b^{y]} \rangle,\label{eq:Delta_1}\\
    \Delta_{2,\mathbf{\delta}_n}=\frac{2}{L_x L_y}\sum_{\langle\langle i,j \rangle\rangle}\sum_{\langle c,d\rangle_{+\mathbf{\delta}_n}} \langle {S}_c^{[x}{S}_d^{y]} \rangle,\label{eq:Delta_2}
\end{gather}
whereby $\langle a,b\rangle_{-\mathbf{\delta}_n}$ are bonds in the four rhombuses $(a,b,i,j)$ that have a negative sign for the DM interaction along the bond $a\rightarrow b$ (see Fig.~\ref{fig:Fig-1} (c)). An average is taken over all such rhombuses in the MPS unit cell. An example of one such rhombus with the bond $\langle a,b\rangle_{-\mathbf{\delta}_1}$ is shown in Fig.~\ref{fig:MFT-lattice} (b). In Eq.~\eqref{eq:Delta_2}, $\langle c,d\rangle_{\mathbf{\delta}_n}$ are nearest neighbour bonds oriented along $\mathbf{\delta}_n$, such that $(c,j,d,i)$ forms a rhombus. An example of such a rhombus with the bond $\langle c,d\rangle_{\mathbf{\delta}_1}$ is shown in Fig.~\ref{fig:MFT-lattice} (c). Finally, we average over all bond directions $\mathbf{\delta}_n$ in the triangular lattice.
\begin{gather}
    \Delta_1=\frac{1}{3}\left(  \Delta_{1,\mathbf{\delta}_1}+\Delta_{1,\mathbf{\delta}_2}+\Delta_{1,\mathbf{\delta}_3} \right),\\
    \Delta_2=\frac{1}{3}\left(  \Delta_{2,\mathbf{\delta}_1}+\Delta_{2,\mathbf{\delta}_2}+\Delta_{2,\mathbf{\delta}_3} \right),
\end{gather} 
The effect of $\Delta_1$ and $\Delta_2$ is to renormalize the nearest and second nearest neighbour Heisenberg interactions into an effective XXZ type Hamiltonian for states along the $D_z=0$ axis. The heuristic Hamiltonian can be written as
\begin{gather}
    H_{\text{eff,XXZ}}=\sum_{\langle i, j\rangle} J_{1}S^z_iS^z_j +\Tilde{J}^{xx/yy}_1 \left(S^x_i S^x_j + S^y_i S^y_j \right) \nonumber\\
    +\sum_{\langle\langle i, j \rangle \rangle}  \Tilde{J}^{zz}_2S^z_iS^z_j +J_{2}  \left(S^x_i S^x_j + S^y_i S^y_j \right)\nonumber\\+\sum_{\langle\langle\langle i,j \rangle\rangle\rangle} J_3 \mathbf{S}_i \cdot \mathbf{S}_j
    +H_{\text{ring}},\label{eq:H_MFT_VBS}
\end{gather}
where we define $\Tilde{J}^{xx/yy}_1=J_1+\Delta_1$ and $\Tilde{J}^{zz}_2=J_2+\Delta_2$. 

We show in Fig.~\ref{fig:MFT-VBS}, the values of these renormalized spin interactions along the $D_z=0$ axis of the phase diagram in Fig.~\ref{fig:Fig-3_VBS_phase_diagram} (a). We find that $J_r/\Tilde{J}_1^{xx/yy}$ decreases and $\Tilde{J}_2^{zz}/J_1$ increases with increasing $J_{r_1}/J_r$, when starting from the VBS phase. 
Based on qualitatively similar trends from \cite{Cookmeyer_four_spin,schultz2023electric}, we expect a CSL, followed by a $120^{\circ}$ order, to appear with increasing $J_{r_1}/J_r$. In agreement with our prediction, we find from the phase diagram in Fig.~\ref{fig:Fig-3_VBS_phase_diagram} (a) (along $D_z=0$) that a CSL phase is favoured with increasing $J_{r_1}/J_r$, followed by a $120^{\circ}$ order for even larger values.
Hence, this heuristic model can explain the appearance of the CSL with increasing $J_{r_1}$ when starting from a VBS phase. Even though our heuristic spin model in Eq.~\eqref{eq:H_MFT_VBS} can qualitatively describe the correct trend, it is of the XXZ type and therefore has a $U(1)$ symmetry, and not the full $SU(2)$ symmetry of the $J_1$-$J_{\text{ring}}$ model used in \cite{Cookmeyer_four_spin,schultz2023electric}.

\section{Discussion}\label{sec:disc}

In this study, we have derived an effective spin model that describes the low-energy physics of the triangular lattice Hubbard model close to the Mott transition in the presence of a weak SOC. Apart from the Heisenberg interactions and conventional ring exchange coupling, we found that the SOC generates DM interactions, anisotropic symmetric exchange interactions, and SOC-mediated ring exchange interactions. The resulting model is quite complicated, so we have restricted attention to the spin interactions that are leading order in $v_z/t$. In particular, of the SOC-induced terms, we keep only the nearest neighbour DM interaction $D_z$ and a SOC-mediated ring exchange interaction $J_{r_1}$. We have treated the strengths of these interactions as independent parameters for a clear understanding of the role of each interaction. 

We have used iDMRG to map out the quantum phase diagrams for this simplified spin model in the presence of a weak SOC. We find that the CSL phase ($t/U=0.097$) and the VBS phase ($t/U=0.105$) remain stable along a narrow elongated region as the strength of the SOC is tuned, keeping $D_z/J_{r_1}\sim 0.5$. In the region of stability, the effect of the DM interaction $D_z$, compensates for the effect of the SOC-mediated ring exchange interaction $J_{r_1}$. Both these interactions individually prefer a magnetically ordered $120^{\circ}$ state, but with opposite handedness. We provided a qualitative argument to support this stabilization mechanism. Further, using a heuristic model we explained the eventual disappearance of the CSL phase along the line of stability $D_z/J_{r_1}\sim 0.5$. This heuristic model can predict the extent of the CSL phase quite accurately compared to the exact iDMRG results. Additionally, from our iDMRG results, we observed that, starting from a VBS phase, increasing the strength of the $J_{r_1}$ can drive the system into the CSL phase. We construct a different heuristic model to explain this in terms of renormalized spin interactions. The qualitative arguments and heuristic models are consistent with the iDMRG phase diagrams, and provide a physical understanding of their structure.

In this study, we have assumed a weak SOC, that is $v_z/t<1$. This justified our choice to neglect the anisotropic symmetric exchange $\Gamma_1$, in the regime where $\Gamma_1<J_{r_1}\implies v_z/t<20\left(t/U \right)^2$. However, this approximation breaks down beyond $J_{r_1}/J_r\approx 0.4$. We expect that a sufficiently large $\Gamma_1$ destabilizes the CSL and VBS phases, and therefore the extent over which the CSL (and VBS) phase is stabilized in Figs.~\ref{fig:Fig-2_CSL_phase_diagram} (and \ref{fig:Fig-3_VBS_phase_diagram}) may become smaller. We have chosen $\mathbf{v}$ is along $\hat{\mathbf{z}}$ for simplicity. It would be interesting to investigate if the same stabilization mechanism discussed in this paper is applicable if $\mathbf{v}$ had non-zero components along $\hat{\mathbf{x}}$ and $\hat{\mathbf{y}}$.

In our model, we have treated the strengths of the DM interaction $D_z$, and the SOC-mediated ring exchange term $J_{r_1}$, as independent parameters to investigate the effect of each contribution more clearly. In the single-band Hubbard model, these parameters are characterized by a single parameter $v_z/t$ in the strong coupling expansion and are related (see Appendix \ref{sec:full-model}). However, it is conceivable that similar types of SOC-mediated interactions appear in the weak Mott insulating regime of a more complex multi-band model, leading to exchange coupling strengths parameterized by multiple independent microscopic energy scales.

In conclusion, we have demonstrated that it is possible to realize novel quantum ground states, such as the CSL and the VBS, that are stabilized by a SOC in the triangular lattice Hubbard model in the weak Mott regime. It would be interesting to explore if the same stabilization mechanism applies to other types of lattices, complicated multi-band models, 3D systems, or for other types of quantum spin liquids. It is also potentially interesting to investigate the effect of different kinds of SOCs. Our results might serve as a motivation for experimental research into the realization of exotic quantum ground states in weak-Mott insulators with a SOC.

\begin{acknowledgements}
This work was supported by 
the Natural Science and Engineering Council of Canada (NSERC) Discovery Grant No. RGPIN-2023-03296 and the Center for Quantum Materials at the University of Toronto. A.M. is supported by the Lachlan Gilchrist Fellowship from the University of Toronto. D.S. is supported by the Ontario Graduate Scholarship. Computations were performed on the Cedar cluster hosted by WestGrid and SciNet in partnership with the Digital Research Alliance of Canada.
\end{acknowledgements}

\onecolumngrid
\diamondrule

\appendix
\section{Details of the Canonical Transformation}\label{sec:can-trans}
In this section, we derive the effective low-energy spin Hamiltonian of the spin-orbit coupled Hubbard model in the weak Mott insulating regime, by performing a strong coupling expansion \cite{t_U_Hubbard_MacDonald,SW_transformation}. We start from the Hubbard model in the presence of a SOC defined in Eq.~\eqref{eq:Hubbard model with SOI}. To incorporate the effect of the SOC in a simple form, we use a modified hopping $\Tilde{t}$ (defined in Eq.~\eqref{eq:Hopping with SOI}) \cite{Spin_triplet_VBS,Krempa_2012} that mediates the SOC. Since we are interested in the half-filled limit, the portion of the Hamiltonian that can induce high-energy fluctuations are those that may change the number of doubly occupied sites. We therefore split the kinetic term into three parts: $T^+$ increases the number of doubly occupied sites, $T^-$ decreases the number of doubly occupied sites, and $T^0$ does not change the number of doubly occupied sites.
\begin{gather}
H=T^{+}+T^{-}+T^{0}+V,\label{eq:H_break}
\end{gather}
where we defined
\begin{align}
T^+ =& -\sum_{i,j}\sum_{\alpha\beta} \Tilde{t}_{ij,\alpha\beta} n_{i,-\alpha}c^\dagger_{i\alpha}c_{j\beta}h_{j,-\beta}, \label{eq:T+}\\
T^- =& -\sum_{i,j}\sum_{\alpha\beta} \Tilde{t}_{ij,\alpha\beta} h_{i,-\alpha} c^\dagger_{i\alpha}c_{j\beta} n_{j,-\beta} ,\label{eq:T-}\\
T^0 =& -\sum_{i,j}\sum_{\alpha\beta} \Tilde{t}_{ij,\alpha\beta}\left(h_{i,-\alpha}c^\dagger_{i\alpha}c_{j\beta}h_{j,-\beta} + n_{i,-\alpha}c^\dagger_{i\alpha}c_{j\beta}n_{j,-\beta}\right), \label{eq:T0}\\
V=& \sum_i U n_{i\uparrow} n_{i\downarrow},\label{eq:V}
\end{align}
Here, $h_{i,\sigma} = 1-n_{i,\sigma}$ is the projector onto empty sites, $n_{i,\sigma}$ is the projector onto occupied sites. We perform a Schrieffer-Wolff transformation \cite{SW_transformation} to find the effective low-energy Hamiltonian at half-filling in the regime $\Tilde{t}/U\ll 1$. This canonical transformation, generated by $S$, eliminates charge fluctuations to the high-energy doubly-occupied sector. The effective Hamiltonian is of the form: 
\begin{gather}
    H_{\text{eff}}=e^{iS}H e^{-iS},
\end{gather}
and this expression can be expanded using the Hausdorff-Baker-Campbell formula as:
\begin{gather}
    H_{\text{eff}}=H+i\left[S,H \right]+\frac{i^2}{2}\left[S,\left[ S, H \right]\right]+\cdots.
\end{gather}
$S$ is calculated order by order in $\Tilde{t}/U$ so that fluctuations to the doubly occupied sector are eliminated at each order in $\Tilde{t}/U$ . We can expand the generator $S$, as
$S=S^{(1)}+S^{(2)}+S^{(3)}+\cdots$, where $S^{n}\propto \left( \Tilde{t}/U \right)^n$. To compute ring-exchange type terms we need to calculate $H_{\text{eff}}$ to $\mathcal{O}(\Tilde{t}^4/U^3)$. Therefore we require the expansion of the generator $S$ to $\mathcal{O}(\Tilde{t}^3/U^3)$ or up to  $S^{(3)}$. We find that the first three terms of the expansion are:
\begin{align}
    iS^{(1)} = &\frac{1}{U}(T^+-T^-), \label{eq:S1} \\
    iS^{(2)} = &\frac{1}{U^2}\left([T^+, T^0]+[T^-, T^0] \right),\label{eq:S2} \\
    iS^{(3)} = &\frac{1}{U^3}([[T^+,T^0],T^0] - [[T^-,T^0],T^0])\nonumber\\ 
    &+\frac{1}{4U^3}([[T^+,T^0],T^+] - [[T^-,T^0],T^-]) \nonumber\\ 
    &+\frac{2}{3U^3}([T^+,[T^+,T^-]] + [T^-,[T^+,T^-]]).\label{eq:S3}
\end{align}

Using these the generators in Eqs.~\eqref{eq:S1}, \eqref{eq:S2}, \eqref{eq:S3},  we calculate $H_{\text{eff}}$ to $\mathcal{O}(\Tilde{t}^4/U^3)$. We find that
\begin{gather}
H_\text{eff} = -\frac{1}{U}T^- T^+ + \frac{1}{U^2}T^-T^0T^+ +\frac{1}{U^3}\left( T^- T^+ T^- T^+ -T^- T^0 T^0 T^+ - \frac{1}{2} T^- T^- T^+ T^+ \right),\label{eq:H_eff_app}
\end{gather}
where the $T^{+/-/0}$ are as defined in Eqs.~\eqref{eq:T+}, \eqref{eq:T-}, \eqref{eq:T0}. We observe that to a given order in $\Tilde{t}/U$, $H_{\text{eff}}$ has an equal number of $T^+$ and $T^-$ operators, and therefore no term can take it out of the singly-occupied (half-filled) sector. Since the site indices are not specified in Eq.\eqref{eq:H_eff_app}, we take into account the contributions from all possible exchange pathways generated by $H_{\text{eff}}$ that start from a half-filled background and take it back to a half-filled background. Next, we convert the chain of second-quantized fermionic operators into the spin operators acting in the half-filled subspace. We use a custom made code to perform these two steps. For simplicity, we choose the nearest neighbour hoppings to be uniform $t_{ij}=t$. We also choose $\mathbf{v}_{ij}$ to be uniform and pointing along $\hat{\mathbf{z}}$, $\mathbf{v}_{ij}=v_{z} \hat{\mathbf{z}}$ (conventions for the positive directions is defined in Fig.~\ref{fig:Fig-1} (c)). With these simplifying choices, we put the spin Hamiltonian on the triangular lattice. In doing so, we consider all possible permutations of the site indices placed on the triangular lattice, and all possible orientations of $\mathbf{v}_{ij}$ in the lattice. This finally gives us the spin Hamiltonian on the triangular lattice, described in Appendix \ref{sec:full-model}.

\section{Full model with Spin-Orbit interactions}\label{sec:full-model}

The effective spin Hamiltonian obtained from the strong coupling expansion of the Hubbard model has Heisenberg interactions, the conventional ring exchange interaction, DM interactions, anisotropic symmetric exchange interactions,  and SOC-mediated ring exchange interactions. The full Hamiltonian including all these interactions is
\begin{align}
    H=&\sum_{\langle ij \rangle}J_1 \mathbf{S}_i\cdot \mathbf{S}_j
     +\sum_{\langle\langle ij \rangle \rangle}J_2 \mathbf{S}_i\cdot \mathbf{S}_j
     +\sum_{\langle \langle\langle ij \rangle \rangle \rangle}J_3 \mathbf{S}_i\cdot \mathbf{S}_j\\
     +&\sum_{\langle ij \rangle} D^{z}_1 {S}_i^{[x}{S}_j^{y]}
     +\sum_{\langle \langle ij \rangle \rangle} D^{z}_2 {S}_i^{[x}{S}_j^{y]}
     +\sum_{\langle \langle \langle ij \rangle \rangle \rangle} D^{z}_3 {S}_i^{[x}{S}_j^{y]}\\
     +&\sum_{\langle ij \rangle} S_i^a \Gamma^{ab}_{1} S_j^b
     +\sum_{\langle \langle ij \rangle \rangle} S_i^a \Gamma^{ab}_{2} S_j^b
     +\sum_{\langle \langle \langle ij \rangle \rangle \rangle} S_i^a \Gamma^{ab}_{3} S_j^b \\
     +&\sum_{(ijkl) \in R}J_r  \left( \left( \mathbf{S}_i\cdot \mathbf{S}_j \right) \left( \mathbf{S}_k\cdot \mathbf{S}_l \right)
     +\left( \mathbf{S}_j\cdot \mathbf{S}_k \right) \left( \mathbf{S}_l\cdot \mathbf{S}_i \right)
     -\left( \mathbf{S}_i\cdot \mathbf{S}_k \right) \left( \mathbf{S}_j\cdot \mathbf{S}_l \right) \right)\\
     +&\sum_{(ijkl) \in R}J_{r_1}\left( S_i^{[x}S_j^{y]}\left(S_k^xS_l^x+S_k^yS_l^y\right)+S_l^{[x}S_i^{y]}\left(S_j^xS_k^x+S_j^yS_k^y\right) \right)+\sum_{\langle ijkl \rangle}2J_{r_1}\left(S_j^{[x}S_l^{y]}S_i^{z} S_k^{z} \right)\\
     -&\sum_{(ijkl) \in R}J_{r_1}\left( S_j^{[x}S_k^{y]}\left(S_l^xS_i^x+S_l^yS_i^y\right)+S_k^{[x}S_l^{y]}\left(S_i^xS_j^x+S_i^yS_j^y\right) \right)\\
     +&\sum_{(ijkl) \in R}J_{r_2}\left(\left(\mathbf{S}_i \overset{z}{\cdot}\mathbf{S}_j\right)\left(\mathbf{S}_k \overset{z}{\cdot}\mathbf{S}_l\right)+
    \left(\mathbf{S}_j \overset{z}{\cdot}\mathbf{S}_k\right)\left(\mathbf{S}_i \overset{z}{\cdot}\mathbf{S}_l\right)-2\left(\mathbf{S}_i \overset{z}{\cdot}\mathbf{S}_k\right)\left(\mathbf{S}_j \overset{z}{\cdot}\mathbf{S}_l\right)\right)\\
    +&\sum_{(ijkl) \in R}J_{r_2}\left(\left(\mathbf{S}_i \cdot\mathbf{S}_j\right)\left(\mathbf{S}_k \cdot\mathbf{S}_l\right)+
    \left(\mathbf{S}_j \cdot\mathbf{S}_k\right)\left(\mathbf{S}_i \cdot\mathbf{S}_l\right)
    +2\left(\mathbf{S}_i \overset{z}{\cdot}\mathbf{S}_k\right)\left(\mathbf{S}_j \cdot\mathbf{S}_l\right)
    -2\left(\mathbf{S}_i \cdot\mathbf{S}_k\right)\left(\mathbf{S}_j \overset{z}{\cdot}\mathbf{S}_l\right)   
    \right)\\
     +&\sum_{(ijkl) \in R}J_{r_3}\left(S_i^{[x}S_j^{y]}S_k^zS_l^z-S_j^{[x}S_k^{y]}S_l^zS_i^z-S_k^{[x}S_l^{y]}S_i^zS_j^z+S_l^{[x}S_i^{y]}S_j^zS_k^z-2S_j^{[x}S_l^{y]}S_i^{z} S_k^{z} 
    \right)\\
     +&\sum_{\langle ijkl \rangle}J_{r_4} \left(\left(\mathbf{S}_i \overset{z}{\cdot}\mathbf{S}_j\right)\left(\mathbf{S}_k \overset{z}{\cdot}\mathbf{S}_l\right)+
    \left(\mathbf{S}_j \overset{z}{\cdot}\mathbf{S}_k\right)\left(\mathbf{S}_i \overset{z}{\cdot}\mathbf{S}_l\right)- \left(\mathbf{S}_i \cdot\mathbf{S}_k\right)\left(\mathbf{S}_j \cdot\mathbf{S}_l\right)\right),
\end{align}
where the sum $(ijkl) \in R$ is over the three types of rhombuses in a triangular lattice, and $i$ is a sharp vertex of the rhombus $(i,j,k,l)$ counted anti-clockwise (see Fig.~\ref{fig:Fig-1} (a)). We have defined the modified dot product as $\mathbf{S}_i \overset{z}{\cdot} \mathbf{S}_j=S^z_iS^z_j-S^x_iS^x_j-S^y_iS^y_j$. We obtain the strength of the exchange interactions from the strong coupling expansion in terms of the parameters, $t/U$ and $v_z/t$, of the original spin-orbit coupled Hubbard model.
\begin{align}
    & J_1=\left(\frac{4t^2}{U}-\frac{4v_z^2}{3U}-\frac{28t^4}{U^3}-\frac{104 t^2 v_z^2}{3U^3}+\frac{44v_z^4}{3U^3} \right) ,\quad
    && J_2= \frac{4\left(t^2+v_z^2 \right)^2}{U^3},\quad
    && J_3=\left(\frac{4t^4}{U^3}-\frac{40t^2v_z^2}{3U^3}+\frac{4v_z^4}{U^3} \right),\\
    & D^z_1=\left(\frac{8tv_z}{U}-\frac{80t^3v_z}{U^3}-\frac{48tv_z^3}{U^3}\right),\quad
    && D^z_2=0,\quad
    && D^z_3=\left(\frac{16t^3v_z}{U^3}-\frac{16tv_z^3}{U^3} \right),\\
    & \Gamma^{ab}_1=\left(\frac{8v_z^2}{U}-\frac{32t^2v_z^2}{U^3}-\frac{64 v_z^4}{U^3} \right)\left(\delta_{az}\delta_{bz}-\frac{\delta_{ab}}{3} \right),\quad
    && \Gamma^{ab}_2=0,\quad
    && \Gamma^{ab}_3=\frac{32t^2v_z^2}{U^3}\left(\delta_{az}\delta_{bz}-\frac{\delta_{ab}}{3} \right),
\end{align}
\begin{align}
    J_r=\frac{80t^4}{U^3},\quad \quad \quad
    J_{r_1}=\frac{160t^3 v_z}{U^3},\quad \quad \quad
    J_{r_2}=\frac{80t^2 v_z^2}{U^3},\quad \quad \quad
    J_{r_3}=\frac{160t v_z^3}{U^3},\quad \quad \quad
    J_{r_4}=\frac{80v_z^4}{U^3}.
\end{align}
Henceforth, we work in the limit of a small SOC, $v_z/t<1$, in addition to $t/U< 1$ and $v_z/U < 1$ used in the strong coupling expansion. 

For simplicity, we construct a simplified spin model by isolating the leading order SOC-mediated interactions as we are working in the limit of a weak SOC. The SOC introduces corrections to the Heisenberg exchanges which are $\mathcal{O}((v_z/t)^2)$, we neglect these corrections. The nearest neighbour DM interaction being the strongest is $\mathcal{O}(t^2/U (v_z/t))$, we neglect all further neighbour DM interactions for simplicity. As we are interested in regions close to the Mott transition we retain the conventional ring exchange $J_r$ which is order $\mathcal{O}(t^4/U^3)$. The leading order SOC-mediated ring exchange $J_{r_1}$ is $\mathcal{O}(t^4/U^3 (v_z/t))$. We retain this term and neglect all higher order SOC-mediated ring exchange terms that are $\mathcal{O}(t^4/U^3 (v_z/t)^n)$ where $n\geq 2$. The most relevant anisotropic symmetric exchange is the nearest neighbour $\Gamma_1$ which is $\mathcal{O}(t^2/U(v_z/t)^2)$. However, for small $v_z/t$, we can neglect $\Gamma_1$ in the regime where $\Gamma_1<J_{r_1}\implies v_z/t<20 \left( t/U \right)^2$. As we are interested in the regime $t/U\sim 0.1$, this approximation of neglecting $\Gamma_1$ is justified for $v_z/t<0.2$. All further neighbour anisotropic symmetric exchanges being $\mathcal{O}(t^4/U^3 (v_z/t)^2)$ are also neglected. Hence, our simplified effective spin model describing the physics of weak Mott insulators in the presence of a weak SOC is
\begin{align}
    H_{\text{eff}}=&\sum_{\langle ij \rangle}J_1 \mathbf{S}_i\cdot \mathbf{S}_j
     +\sum_{\langle\langle ij \rangle \rangle}J_2 \mathbf{S}_i\cdot \mathbf{S}_j
     +\sum_{\langle \langle\langle ij \rangle \rangle \rangle}J_3 \mathbf{S}_i\cdot \mathbf{S}_j
     +\sum_{\langle ij \rangle} D^{z}_1 {S}_i^{[x}{S}_j^{y]}\\
     +&\sum_{\langle ijkl \rangle}J_r  \left( \left( \mathbf{S}_i\cdot \mathbf{S}_j \right) \left( \mathbf{S}_k\cdot \mathbf{S}_l \right)
     +\left( \mathbf{S}_j\cdot \mathbf{S}_k \right) \left( \mathbf{S}_l\cdot \mathbf{S}_i \right)
     -\left( \mathbf{S}_i\cdot \mathbf{S}_k \right) \left( \mathbf{S}_j\cdot \mathbf{S}_l \right) \right)\\
     +&\sum_{\langle ijkl \rangle}J_{r_1}\left( S_i^{[x}S_j^{y]}\left(S_k^xS_l^x+S_k^yS_l^y\right)+S_l^{[x}S_i^{y]}\left(S_j^xS_k^x+S_j^yS_k^y\right) \right)\\
     -&\sum_{\langle ijkl \rangle}J_{r_1}\left( S_j^{[x}S_k^{y]}\left(S_l^xS_i^x+S_l^yS_i^y\right)+S_k^{[x}S_l^{y]}\left(S_i^xS_j^x+S_i^yS_j^y\right) \right)\\
     +&\sum_{\langle ijkl \rangle}2J_{r_1}\left(S_j^{[x}S_l^{y]}S_i^{z} S_k^{z} \right),
\end{align}
where we have parametrized the strengths of the Heisenberg couplings and the conventional ring exchange in terms of $t/U$, as obtained from a strong-coupling expansion. 
\begin{gather}
    J_1=\frac{4t^2}{U}-\frac{28t^4}{U^3},\quad \quad 
    J_2=\frac{4t^4}{U^3},\quad \quad 
    J_3=\frac{4t^4}{U^3},\quad \quad 
    J_r=\frac{80t^4}{U^3}.
\end{gather}
Henceforth, we use $D_z$ to denote the nearest neighbour DM interaction $D^z_1$. Further, in the regime of small $v_z/t$ we treat $D_z$ and $J_{r_1}$ as independent parameters. Finally, this spin model is used in our iDMRG simulations to construct the quantum phase diagrams shown in Figs.~\ref{fig:Fig-2_CSL_phase_diagram} \& \ref{fig:Fig-3_VBS_phase_diagram}, where $t/U=0.097$ and $t/U=0.105$ respectively.

\section{DMRG Calculations}\label{sec:DMRG_cals}

For the iDMRG simulations, we have used the library TeNPy \cite{Hauschild_TeNPy_2018}. We have used a YC6 cylindrical geometry, where one of the edges of the triangular lattice is parallel to the circumference of the cylinder. The circumference was chosen to have a length $L_y=6$ sites in the $\mathbf{a}_1$ direction. In the infinite direction $\mathbf{a}_2$, the MPS unit cell was chosen to have length $L_x=2$. We have used bond dimension $\chi_{\text{max}}=1600$ in all our calculations, and we have focused on the $S^z_{\text{tot}}=0$ sector of the spin model. For the iDMRG simulation, we follow a 3-step approach for each point in the phase diagram. 
\begin{enumerate}
    \item \label{eq:item 1} We initialize the starting state as a product state of up/down spins. Setting the chiral symmetry breaking term to be $J_{\chi}=10^{-5}$, and $\chi_{\text{max}}=20$, we perform 5 iDMRG sweeps. 
    \item \label{eq:item 2} We set $J_{\chi}=0$, and $\chi_{\text{max}}=1600$. Turning the density mixer 'on' to escape from a local minima in the energy landscape, we perform 40 iDMRG sweeps using the output state from step \eqref{eq:item 1} as the initial state.
    \item Using the output state from step \eqref{eq:item 2} as the initial state, we perform iDMRG sweeps with the density mixer 'off' and $\chi_{\text{max}}=1600$, for a maximum of 100 sweeps or until a convergence in energy is achieved to within $\Delta E\sim 10^{-8}$.
\end{enumerate}

Sometimes the iDMRG wavefunction does not converge to the correct ground state when using this 3-step algorithm for some points in the phase diagram. This usually happens if the strength of the ring exchange interactions is large, as it increases frustration by introducing a large number of competing states. In that case, we used the wavefunctions of nearby points in the phase diagram as the initial state for the iDMRG simulation. Finally, we choose the state with the least energy as the true ground state to construct the phase diagrams.

\section{DMRG Data}\label{sec:DMRG_data}
\begin{figure*}
    \centering
    \includegraphics[scale=0.55]{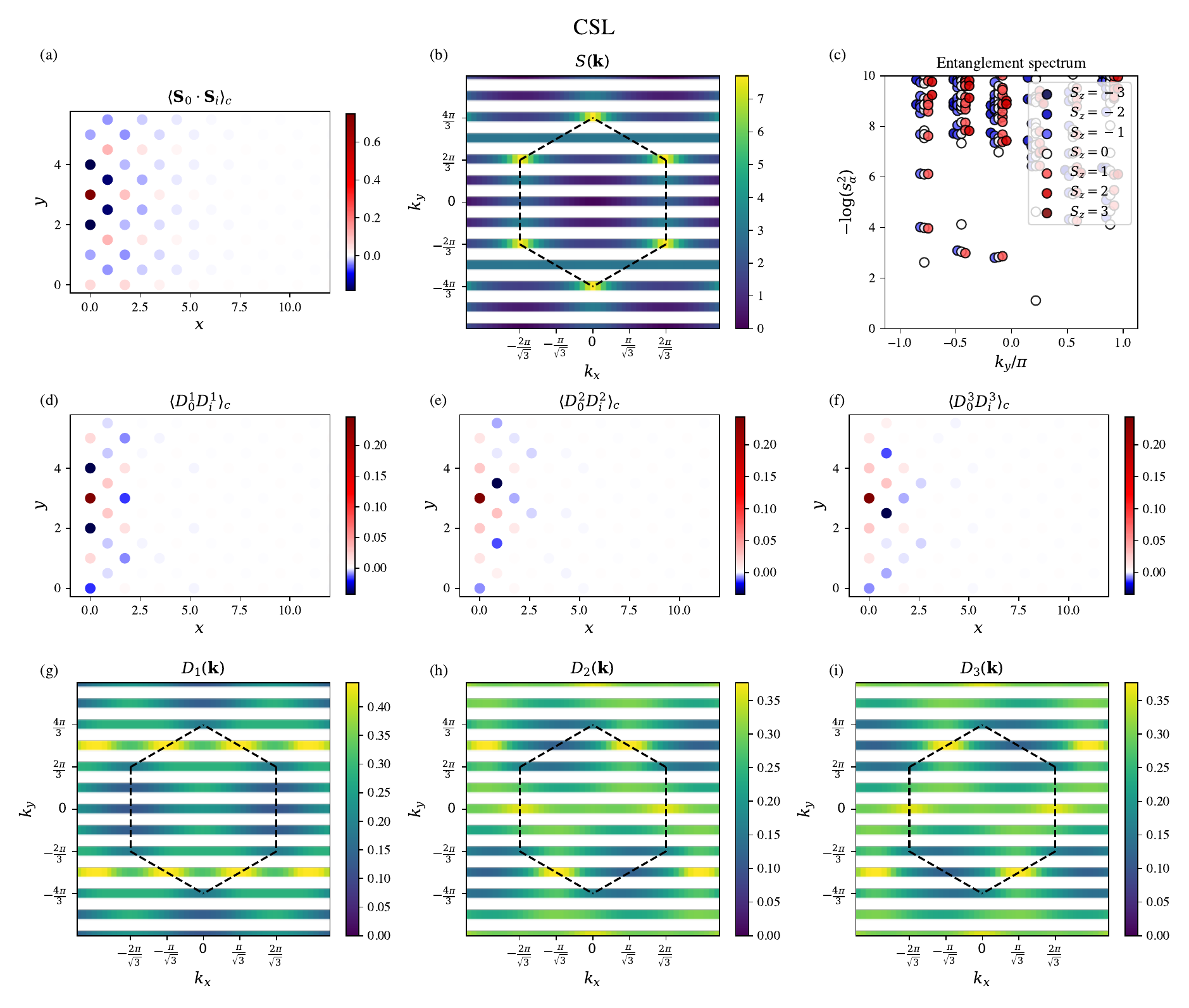}
    \caption{Observables in the CSL phase ($J_1=1$, $J_2=0.01$, $J_3=0.01$, $J_r=0.2$, $D_z/J_1=0.04$, $J_{r_1}/J_r=0.4$). The real space spin correlations $\langle \mathbf{S}_i\cdot \mathbf{S}_j \rangle_c=\langle \mathbf{S}_i\cdot \mathbf{S}_j \rangle-\langle \mathbf{S}_i \rangle \cdot \langle \mathbf{S}_j \rangle$ are shown in (a), the static spin structure factor $S(\mathbf{k})$ in (b). Real space dimer correlations $\left\langle D^n_iD^n_j \right\rangle_c=\left\langle D^n_iD^n_j \right\rangle-\left\langle D^n_i \right\rangle \left\langle D^n_j \right\rangle$ along the the different directions $\mathbf{a}_n$ are shown in (d)-(f), and the static dimer structure factors $D_n(\mathbf{k})$ in (g)-(i). Here we have chosen $\mathbf{R}_0=3 \mathbf{a}_1$. The entanglement spectrum in (c) shows the characteristic structure of the Kalmeyer-Laughlin type CSL \cite{Wen_1991,Li_2008}.}
    \label{fig:CSL}
\end{figure*}

\begin{figure*}
    \centering
    \includegraphics[scale=0.55]{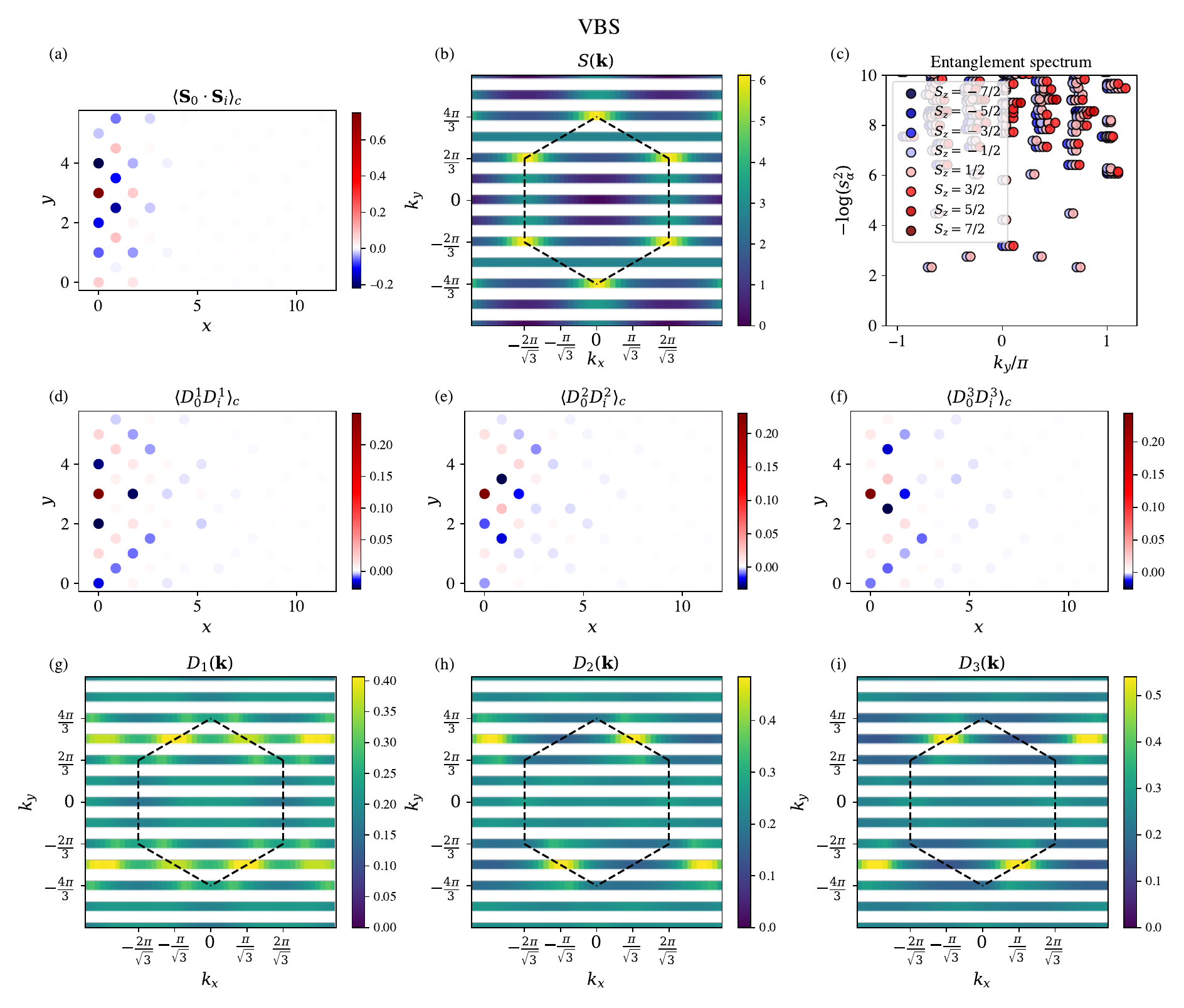}
    \caption{Observables in the VBS phase ($J_1=1$, $J_2=0.012$, $J_3=0.012$, $J_r=0.24$, $D_z=J_{r_1}=0$). The real space spin correlations $\langle \mathbf{S}_i\cdot \mathbf{S}_j \rangle_c=\langle \mathbf{S}_i\cdot \mathbf{S}_j \rangle-\langle \mathbf{S}_i \rangle \cdot \langle \mathbf{S}_j \rangle$ are shown in (a), the static spin structure factor $S(\mathbf{k})$ in (b). Real space dimer correlations $\left\langle D^n_iD^n_j \right\rangle_c=\left\langle D^n_iD^n_j \right\rangle-\left\langle D^n_i \right\rangle \left\langle D^n_j \right\rangle$ along the the different directions $\mathbf{a}_n$ are shown in (d)-(f), and the static dimer structure factors $D_n(\mathbf{k})$ in (g)-(i). Here we have chosen $\mathbf{R}_0=3 \mathbf{a}_1$. Since the peaks of $D_2(\mathbf{k})$ and $D_3(\mathbf{k})$ are the sharpest, the VBS has a preferred bonding axis of the singlets along $\mathbf{a}_2$ and $\mathbf{a}_3$ \cite{schultz2023electric}.}
    \label{fig:VBS}
\end{figure*}

\begin{figure*}
    \centering
    \includegraphics[scale=0.55]{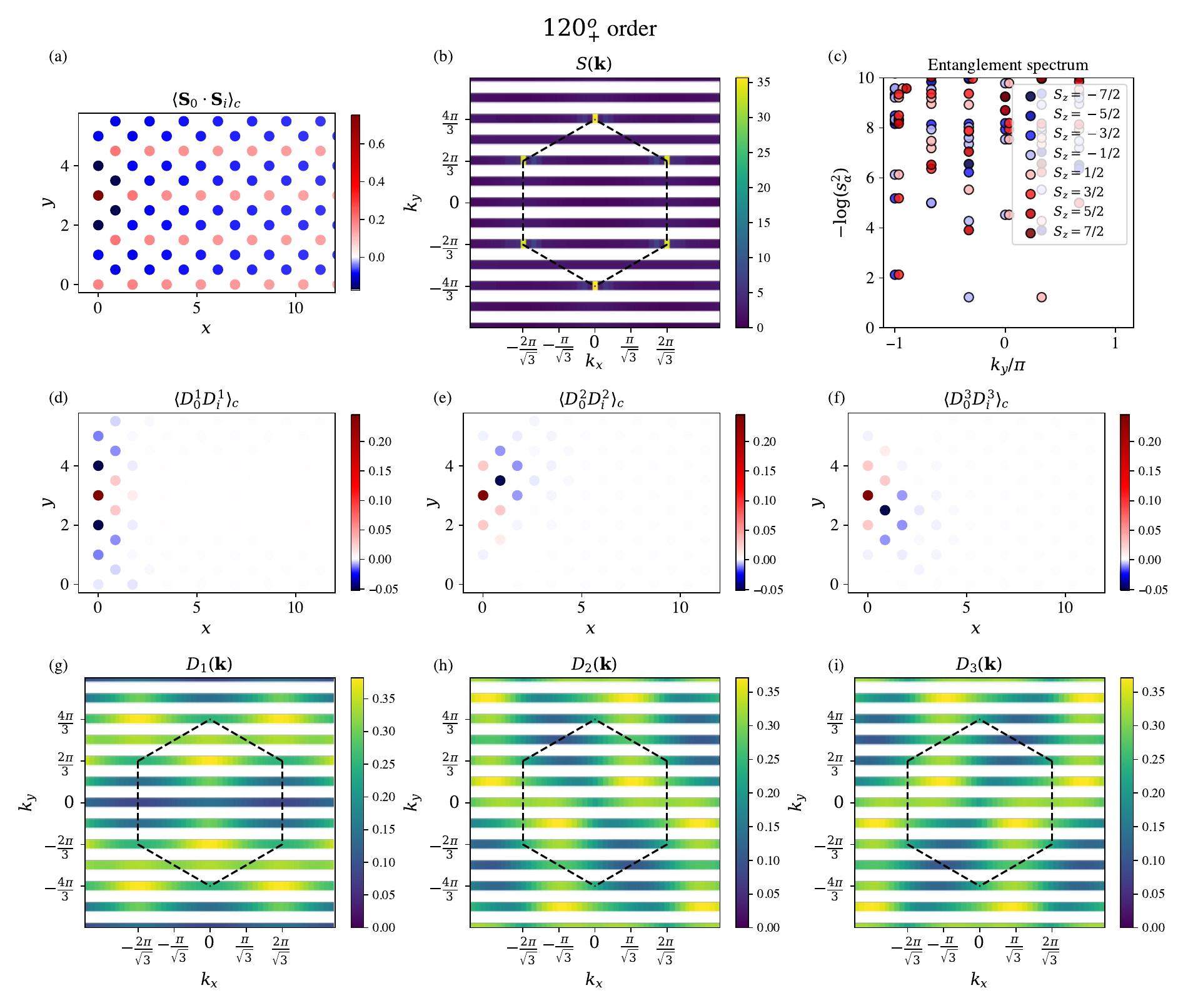}
    \caption{Observables in the $120^{\circ}_{+}$ phase ($J_1=1$, $J_2=0.012$, $J_3=0.012$, $J_r=0.24$, $D_z=0$, $J_{r_1}=0.24$). The real space spin correlations $\langle \mathbf{S}_i\cdot \mathbf{S}_j \rangle_c=\langle \mathbf{S}_i\cdot \mathbf{S}_j \rangle-\langle \mathbf{S}_i \rangle \cdot \langle \mathbf{S}_j \rangle$ are shown in (a), the static spin structure factor $S(\mathbf{k})$ in (b). Real space dimer correlations $\left\langle D^n_iD^n_j \right\rangle_c=\left\langle D^n_iD^n_j \right\rangle-\left\langle D^n_i \right\rangle \left\langle D^n_j \right\rangle$ along the the different directions $\mathbf{a}_n$ are shown in (d)-(f), and the static dimer structure factors $D_n(\mathbf{k})$ in (g)-(i). Here we have chosen $\mathbf{R}_0=3 \mathbf{a}_1$. The spin correlations in (a) are long-ranged, therefore this is an ordered phase. All the correlations in this phase are similar to the $120^{\circ}_{-}$ phase in Fig.~\ref{fig:120m}, except for the value of the handedness $\omega$.}
    \label{fig:120p}
\end{figure*}

\begin{figure*}
    \centering
    \includegraphics[scale=0.55]{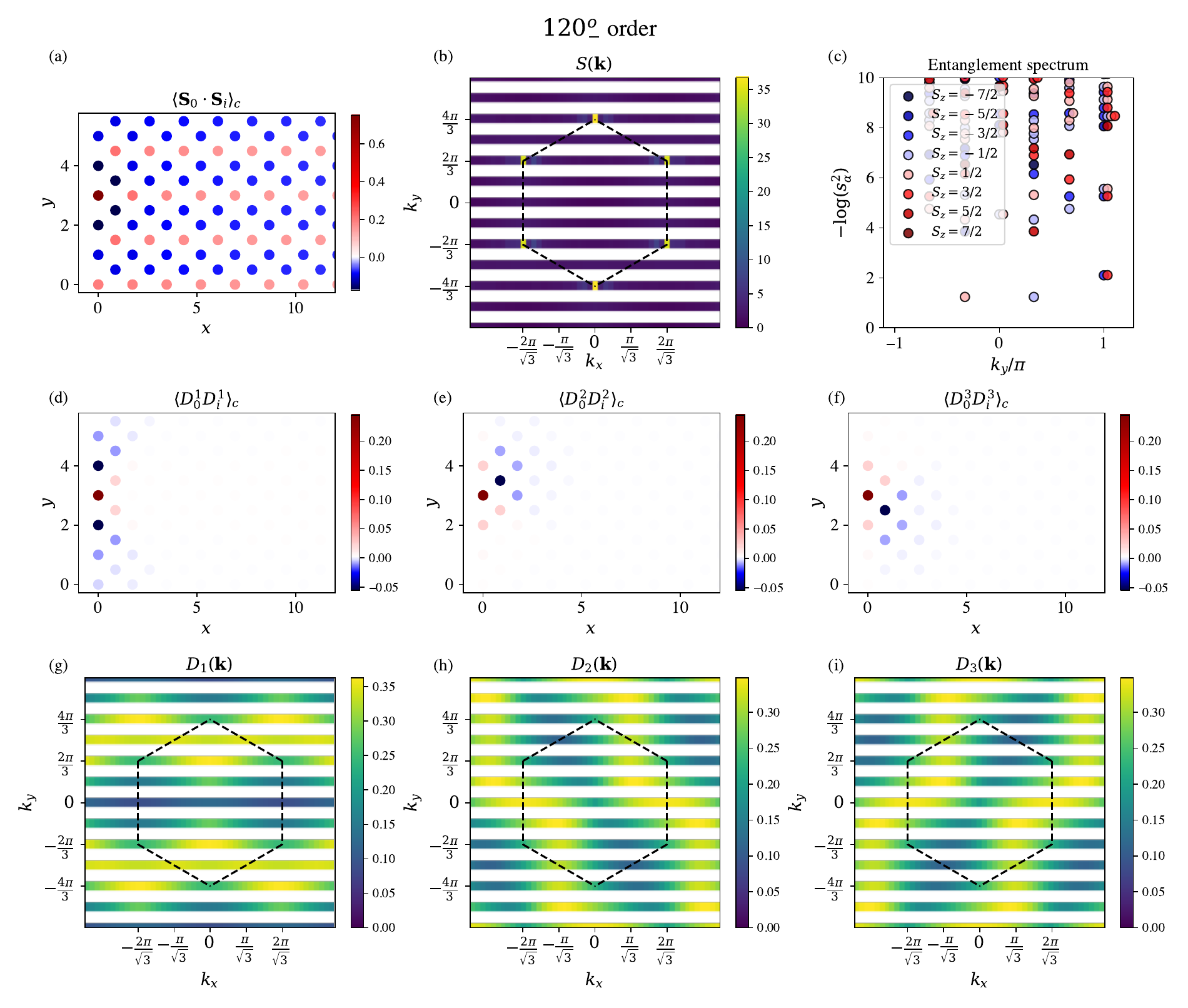}
    \caption{Observables in the $120^{\circ}_{-}$ phase ($J_1=1$, $J_2=0.012$, $J_3=0.012$, $J_r=0.24$, $D_z=0.1$, $J_{r_1}=0$). The real space spin correlations $\langle \mathbf{S}_i\cdot \mathbf{S}_j \rangle_c=\langle \mathbf{S}_i\cdot \mathbf{S}_j \rangle-\langle \mathbf{S}_i \rangle \cdot \langle \mathbf{S}_j \rangle$ are shown in (a), the static spin structure factor $S(\mathbf{k})$ in (b). Real space dimer correlations $\left\langle D^n_iD^n_j \right\rangle_c=\left\langle D^n_iD^n_j \right\rangle-\left\langle D^n_i \right\rangle \left\langle D^n_j \right\rangle$ along the the different directions $\mathbf{a}_n$ are shown in (d)-(f), and the static dimer structure factors $D_n(\mathbf{k})$ in (g)-(i). Here we have chosen $\mathbf{R}_0=3 \mathbf{a}_1$. The spin correlations in (a) are long-ranged, therefore this is an ordered phase. All the correlations in this phase are similar to the $120^{\circ}_{+}$ phase in Fig.~\ref{fig:120p}, except for the value of the handedness $\omega$.}
    \label{fig:120m}
\end{figure*}

\begin{figure*}
    \centering
    \includegraphics[scale=0.55]{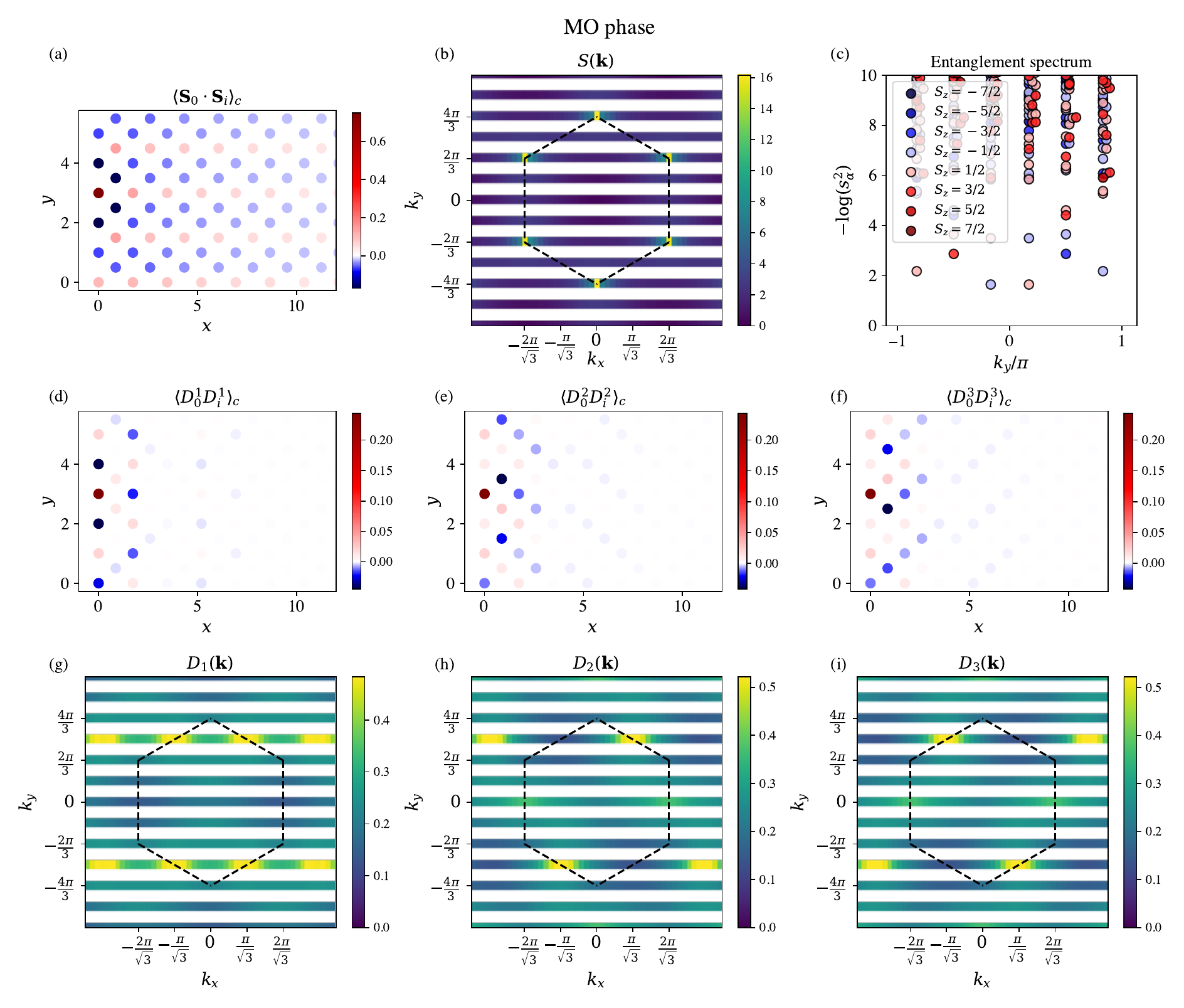}
    \caption{Observables in the magnetic ordered (MO) phase ($J_1=1$, $J_2=0.012$, $J_3=0.012$, $J_r=0.24$, $D_z=0.065$, $J_{r_1}=0.072$). The real space spin correlations $\langle \mathbf{S}_i\cdot \mathbf{S}_j \rangle_c=\langle \mathbf{S}_i\cdot \mathbf{S}_j \rangle-\langle \mathbf{S}_i \rangle \cdot \langle \mathbf{S}_j \rangle$ are shown in (a), the static spin structure factor $S(\mathbf{k})$ in (b). Real space dimer correlations $\left\langle D^n_iD^n_j \right\rangle_c=\left\langle D^n_iD^n_j \right\rangle-\left\langle D^n_i \right\rangle \left\langle D^n_j \right\rangle$ along the the different directions $\mathbf{a}_n$ are shown in (d)-(f), and the static dimer structure factors $D_n(\mathbf{k})$ in (g)-(i). Here we have chosen $\mathbf{R}_0=3 \mathbf{a}_1$. Even though the spin correlations in (a) are long-ranged, they are shorter-ranged than the $120^{\circ}_{\pm}$ phases in Figs.~\ref{fig:120p}, \ref{fig:120m}.}
    \label{fig:MO}
\end{figure*}

Using iDMRG we obtain a set of useful data that is used to characterize the nature of the quantum ground state. Firstly, iDMRG gives the ground state energy per lattice site. In addition to local magnetization $\langle S^z_i \rangle$, we obtain long-range real space spin correlations, $\langle \mathbf{S}_i \cdot \mathbf{S}_j\rangle$, along the infinite direction $x$ of the cylinder. From the real space spin correlations, we can calculate the static spin structure factor \eqref{eq:STF}.
We also obtain real space long-ranged dimer correlations of the spins, $\left\langle D^n_iD^n_j \right\rangle$, where $D^n_i=\mathbf{S}_i\cdot \mathbf{S}_{i+\mathbf{a}_n}$ is the dimer operator along the direction $\mathbf{a}_n$ (see Fig.~\ref{fig:Fig-1} (a)). Using this the static dimer structure factor \eqref{eq:DSF} can be calculated.
Further, we also obtain the scalar chirality $\chi$ as defined in \eqref{eq:chi}, and the handedness $\omega$ as defined in \eqref{eq:handedness}. Most importantly, from iDMRG we can obtain the momentum-resolved entanglement spectrum. Figs.~\ref{fig:CSL}-\ref{fig:MO} show the various correlation functions and entanglement spectra for all the different phases appearing in Figs.~\ref{fig:Fig-2_CSL_phase_diagram} \& \ref{fig:Fig-3_VBS_phase_diagram}.

\begin{figure}
    \centering
    \includegraphics{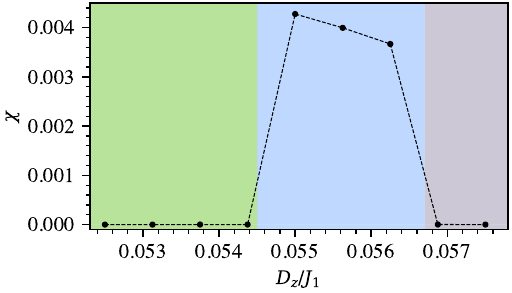}
    \caption{A zoom-in of Fig.~\ref{fig:Fig-3_VBS_phase_diagram} (b) ($t/U=0.105$, $J_{r_1}/J_r=0.3$) showing clearly the phase boundaries of the CSL. The background colors denote the various phases as in Fig.~\ref{fig:Fig-3_VBS_phase_diagram}.}
    \label{fig:zoom}
\end{figure}

\section{Different types of $120^{\circ}$ order}\label{sec:120pm_details}

\begin{figure*}
    \centering
    \hspace{-0.8cm}
    \includegraphics{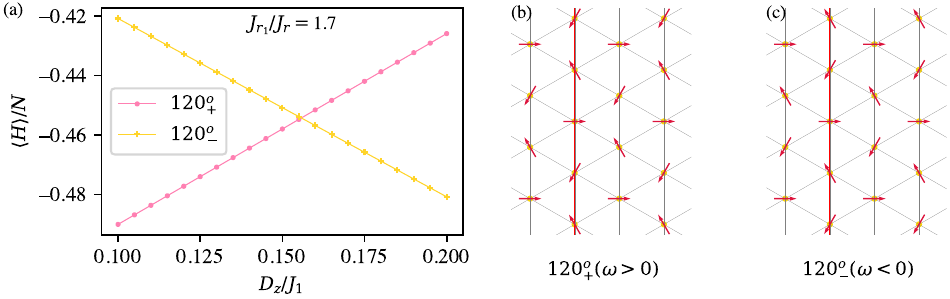}
    \caption{The expectation value of the Hamiltonian, $\braket{\psi_a|H|\psi_a}$ as a function of $D_z/J_1$, computed using representative states $\ket{\psi_a}$ from the $120^{\circ}_{+}$ and $120^{\circ}_{-}$ phases (see Appendix \ref{sec:120pm_details} for details) along the $J_{r_1}/J_r=1.7$ axis in Fig.~\ref{fig:Fig-2_CSL_phase_diagram} (a). The crossing in the value of $\langle H \rangle$ between the $120^{\circ}_{+}$ and $120^{\circ}_{-}$ phases is at $D_z/J_1\approx 0.155$. This value is consistent with the location of the phase boundary in Fig.~\ref{fig:Fig-2_CSL_phase_diagram} (a), between these phases. The two types of oppositely handed $120^{\circ}$ order are visualized using classical spins in (b) and (c) for the $120^{\circ}_+$ ($\omega>0$) and $120^{\circ}_-$ ($\omega<0$) phases respectively. Along the red lines from bottom to top, (which is also a positive direction of $\mathbf{v}$) the spins in (b) and (c) wind anti-clockwise and clockwise respectively, indicating their handedness $\omega$.
    }
    \label{fig:level_crossing}
\end{figure*}

In this study we have identified two different types of $120^{\circ}$ ordered phases, the $120^{\circ}_+$ and $120^{\circ}_-$ phase. These are identified based on their handedness $\omega$ defined in Eq.~\eqref{eq:handedness}. The $120^{\circ}_+$ has $\omega>0$, and $120^{\circ}_-$ has $\omega<0$.
A visualization of these two phases using classical spins is shown in Fig.~\ref{fig:level_crossing} (b)-(c). Previous iDMRG studies \cite{Cookmeyer_four_spin,schultz2023electric} have identified the $120^{\circ}$ phase without subdividing it into the $120^{\circ}_{\pm}$ phases. We have checked that the $120^{\circ}$ phase appearing in \cite{Cookmeyer_four_spin,schultz2023electric} has $\omega=0$. A possible explanation for this fact is if the $120^{\circ}$ phase is an equal weight superposition of the $120^{\circ}_{+}$ and $120^{\circ}_{-}$ phases, and in the absence of a SOC the two oppositely handed phases are equally favoured. 

To test this hypothesis, we have computed the overlap in the wavefunctions between representative states of these phases. For the representative state $\ket{\psi_{120^{\circ}}}$ for the $120^{\circ}$ phase, we have used the ground state corresponding to $J_1=1$, $J_2=J_3=0.0$, $J_r=0.16$. For the $\ket{\psi_{120^{\circ}_+}}$ state the corresponding parameters are $J_1=1$, $J_2=J_3=0.01$, $J_r=0.2$, $D_z=0.005$, $J_{r_1}=1.2$, and for the $\ket{\psi_{120^{\circ}_-}}$ state they are  $J_1=1$, $J_2=J_3=0.01$, $J_r=0.2$, $D_z=0.195$, $J_{r_1}=1.2$. We find that $|\langle\psi_{120^{\circ}}|\psi_{120^{\circ}_{+}}\rangle|\sim0.7$ and $|\langle\psi_{120^{\circ}}|\psi_{120^{\circ}_{-}}\rangle|\sim0.7$. Therefore to leading order the $120^{\circ}$ phase is an equal weight superposition of the $120^{\circ}_{+}$ and $120^{\circ}_{-}$ phases. Also we find,  $|\langle\psi_{120^{\circ}_{+}}| \psi_{120^{\circ}_{-}} \rangle| \sim 0.05$, which suggests that there is minimal overlap between the two oppositely handed $120^{\circ}$ phases. 

Further, to distinguish between the $120^{\circ}_+$ and $120^{\circ}_-$ phases, we use these representative states, $\ket{\psi_a}$ ($a\in \{120^{\circ}_+, 120^{\circ}_- \}$), to compute the expectation value of the Hamiltonian parameterized along the axis $J_{r_1}/J_r=1.7$ of the phase diagram in Fig.~\ref{fig:Fig-2_CSL_phase_diagram} (a). We compute $\langle \psi_a|H|\psi_a \rangle$ as a function of $D_z/J_1$ and this is shown in Fig.~\ref{fig:level_crossing}. Even though the representative states are not eigenstates of the Hamiltonian, we observe in Fig.~\ref{fig:level_crossing}, a crossing in the values of $\langle H \rangle$ computed using the $\ket{\psi_{120^{\circ}_+}}$ and $\ket{\psi_{120^{\circ}_-}}$. This suggests that the two differently-handed $120^{\circ}$ phases are preferred on opposite sides of the crossing. The location of this crossing of $\langle H \rangle$ in Fig.~\ref{fig:level_crossing}, also roughly corresponds to the location of the phase boundary between the $120^{\circ}_+$ and $120^{\circ}_-$ (see Fig.~\ref{fig:Fig-2_CSL_phase_diagram} (a)). The crossing in the value of $\langle H \rangle$ computed using representative states is similar to what we typically expect to see in a variational calculation using metastable eigenstates, however in Fig.~\ref{fig:level_crossing} we do not perform a variational calculation. 

Even though we have used particular representative states (corresponding parameters of the Hamiltonian mentioned earlier in this Section) from the different phases to compute the overlap/expectation values, we expect these values to be largely independent of the choice of representative states.

\twocolumngrid

%

\end{document}